\documentclass[a4paper,11pt]{article}
\pdfoutput=1 
\usepackage{jheppub} 
\usepackage{bbm,bm,graphicx,mathtools,color,hyperref,slashed}
\usepackage{amssymb,amsmath}
\usepackage[T1]{fontenc}
\usepackage[caption=false]{subfig}

\graphicspath{{./Figures/}}

\newcommand{\diff}{\mathrm{d}}
\newcommand{\p}{\partial}

\newcommand{\be}{\begin{equation}}      
\newcommand{\ee}{\end{equation}}      
\newcommand{\bea}{\begin{eqnarray}}      
\newcommand{\eea}{\end{eqnarray}}

\newcommand{\im}{\mathrm{i}}

\newcommand{\eref}{\eqref}

\title{Gradient flows without blow-up for Lefschetz thimbles}

\author[a]{Yuya Tanizaki,}
\author[a]{Hiromichi Nishimura,}
\author[b]{and Jacobus J. M. Verbaarschot}

\affiliation[a]{RIKEN BNL Research Center, Brookhaven National Laboratory, Upton, NY 11973 USA}
\affiliation[b]{Department of Physics and Astronomy, Stony Brook University, Stony Brook, NY 11794, USA}

\emailAdd{yuya.tanizaki@riken.jp}
\emailAdd{hnishimura@bnl.gov}
\emailAdd{Jacobus.Verbaarschot@stonybrook.edu}

\abstract{
We propose new gradient flows that define Lefschetz thimbles and do not blow up in a finite flow time. We study analytic properties of these gradient flows, and confirm them by numerical tests in simple examples. 
}

\begin{document}
\maketitle
\section{Introduction}\label{sec:introduction}

To study properties of strongly-correlated many-body systems, numerical simulation provides us powerful tools. Exact diagonalization of the Hamiltonian gives the complete information of physical systems, however it requires an exponentially large amount of the computational cost as the number of particles increases. 
Monte Carlo simulation of the path integral circumvents this problem, and many physical systems of hadron and condensed-matter physics in thermal equilibrium have been successfully studied with this method~\cite{Karsch:2001cy, RevModPhys.67.279, Pollet2012, Yamamoto:2012wy}. 

Monte Carlo simulation is based on importance sampling, and thus the Boltzmann weight $\exp(-S)$ with the classical action $S$ must be positive semi-definite. 
The Boltzmann weight in many interesting systems, however, takes on
complex values, so that the idea of importance sampling cannot be applied, which is called the sign problem ~\cite{PhysRevB.41.9301, Batrouni:1992fj}. 
The conventional solution of this problem is the reweighting method with phase quenching, but this procedure generally revives the exponential complexity and its use is very limited~\cite{Troyer:2004ge}. 
In hadron physics, high-density cold nuclear matter gets a lot of attention because of its relevance to neutron-star physics~\cite{Fukushima:2010bq}, but finite-density quantum chromodynamics (QCD) suffers from the sign problem, and we do not have a technique for \textit{ab initio} computations
of this theory~\cite{Muroya:2003qs}. 

Recently, a new systematic approach to the sign problem has been  developing based on the complexification of path-integral variables, and it is called the Lefschetz-thimble Monte Carlo method~\cite{Cristoforetti:2012su, Cristoforetti:2013wha, Cristoforetti:2014gsa, Aarts:2013fpa, Fujii:2013sra, Mukherjee:2014hsa, Aarts:2014nxa, Tanizaki:2014xba, Cherman:2014sba, Tanizaki:2014tua, Kanazawa:2014qma, Tanizaki:2015pua, DiRenzo:2015foa, Fukushima:2015qza, Tsutsui:2015tua, Tanizaki:2015rda, Fujii:2015bua, Fujii:2015vha, Alexandru:2015xva, Hayata:2015lzj, Alexandru:2015sua, Alexandru:2016gsd, Alexandru:2016ejd, Tanizaki:2016xcu, Fukuma:2017fjq, Alexandru:2017oyw, Nishimura:2017vav}. 
If the integration variables are complexified, there is a great deal  of freedom in the choice of integration contours thanks to Cauchy's theorem. 
Although expectation values of physical observables does not change under the continuous deformation of integration contours,  the strength of the sign problem heavily depends on the choice of contours, so we can expect that there must be an optimal choice. 
For one-dimensional integrals, it is given by the stationary-phase path, and its higher dimensional analogue is called the Lefschetz thimble. 
This  technique has been developed in the context of the hyperasymptotics of multi-dimensional exponential integrals~\cite{pham1983vanishing, Kaminski1994, Howls2271}, and it is now applied in physics not only to the sign problem but also to the resurgence theory~\cite{Witten:2010cx, Witten:2010zr, Harlow:2011ny, Dunne:2012ae,Basar:2013eka,Cherman:2014ofa,
Cherman:2014xia,Dorigoni:2014hea, Felder:2004uy,Marino:2008ya,
Marino:2012zq,Schiappa:2013opa,Misumi:2014bsa, Misumi:2014jua, Behtash:2015kna, Behtash:2015kva, Basar:2015xna, Cherman:2016hcd, Gukov:2016njj, Gukov:2016tnp, Fujimori:2016ljw, Kozcaz:2016wvy, Sulejmanpasic:2016llc}. 
Quite recently, the Lefschetz-thimble method is also discussed in the context of quantum cosmology~\cite{Feldbrugge:2017fcc, DiazDorronsoro:2017hti} through its application to the real-time quantum phenomena~\cite{Tanizaki:2014xba, Cherman:2014sba, Alexandru:2016gsd}. 

In order to find Lefschetz thimbles, all we have to do is to analyze the gradient flow in the space of complexified field configurations $z^i=x^i+\im y^i$~\cite{Witten:2010cx, Witten:2010zr}:
\be
{\diff z^i\over\diff t}=\overline{\left(\p S(z)\over \p z^i\right)}. 
\label{eq:intro_conventional_flow}
\ee
However, as we will see in this paper, solutions of the gradient flow (\ref{eq:intro_conventional_flow}) generically blow up within finite-time intervals, so one must treat them carefully in numerical computations to get the correct answer. 
Instead, we propose a new gradient flow equation, 
\be
{\diff z^i\over \diff t}=\mathrm{e}^{-2\mathrm{Re}(S_B(z))/\Lambda_B}{|D(z)|^2\over |D(z)|^2+\Lambda_F^{-2}}\overline{\left({\p S(z)\over \p z^i}\right)},
\label{eq:intro_new_flow}
\ee
where $S_B$ is the bosonic classical action, $D$ is the fermionic determinant, and the total classical action is $S=S_B-\ln D$. Here, $\Lambda_B$ and $\Lambda_F$ are positive real parameters to be tuned appropriately. 
This new flow equation (\ref{eq:intro_new_flow}) turns out to define the Lefschetz-thimble decomposition of the original integration as the conventional one (\ref{eq:intro_conventional_flow}) does, and all of its solutions do not show blow-up.

This paper is organized as follows: Section~\ref{sec:blowup} gives a brief review on the Lefschetz-thimble method with the conventional gradient flow, and we argue that blow-up generically happens using simple examples. 
In Sec.~\ref{sec:new_flows}, we introduce new gradient flows and justify their use for the Lefschetz-thimble method. Furthermore, we show that blow-up does not occur in that new flow equation. 
In Sec.~\ref{sec:numerics}, we numerically study the gradient flow in simple examples to see its behaviors and check that the sign problem is indeed equally solved compared with the conventional gradient flow.
Concluding remarks are made in Sec.~\ref{sec:conclusion}.  Several technical
details are worked out  in two appendices. In appendix~\ref{app:proof_equivalence}, we give a mathematical proof on the equivalence among gradient flows.  The deviation equation of the gradient flow to compute the Jacobian is given in Appendix~\ref{app:flow_jacobian}, and  in Appendix~\ref{app:complex_str}, we discuss some technical details on the complex structure to claim that our proposal works with gauge symmetries. 

%-------------------------------------------------------------------------------%
\section{Blow-up of conventional gradient flows for Lefschetz thimbles}\label{sec:blowup}
We first review the Lefschetz-thimble approach to the sign problem in Sec.~\ref{sec:review}. In Sec.~\ref{sec:blowup_example}, we explain that the gradient flow conventionally used blows up in a finite time by demonstrating it in simple two examples. 

\subsection{Brief review of Lefschetz-thimble methods for sign problems}\label{sec:review}
Let us consider the following integral, 
\be
Z=\int_{\mathbb{R}^n}\diff^n x \exp\left[-S(x)\right], \label{eq:path_integral}
\ee
where the action $S(x)$ is a complex-valued polynomial in $x$. Since $S(x)$ is complex valued, this integral becomes oscillatory and the sign problem appears. 
The Lefschetz-thimble method is a method to make the sign problem milder by deforming the integration contour $\mathbb{R}^n$ to other $n$-dimensional submanifolds of the complexified space $\mathbb{C}^n$. 

Conventionally, such submanifolds are constructed by solving the following gradient flow~\cite{Cristoforetti:2012su, Cristoforetti:2013wha, Cristoforetti:2014gsa, Aarts:2013fpa, Fujii:2013sra, Mukherjee:2014hsa, Aarts:2014nxa, Tanizaki:2014xba, Cherman:2014sba, Tanizaki:2014tua, Kanazawa:2014qma, Tanizaki:2015pua, DiRenzo:2015foa, Fukushima:2015qza, Tsutsui:2015tua, Tanizaki:2015rda, Fujii:2015bua, Fujii:2015vha, Alexandru:2015xva, Hayata:2015lzj, Alexandru:2015sua, Alexandru:2016gsd, Alexandru:2016ejd, Tanizaki:2016xcu, Fukuma:2017fjq, Alexandru:2017oyw, Nishimura:2017vav}, 
\be
{\diff z^i\over \diff t}=\overline{\left({\p S(z)\over \p z^i}\right)}, 
\label{eq:conventional_flow}
\ee
where $z^i$ is a holomorphic coordinate of $\mathbb{C}^n$: $z^i=x^i+\im y^i$. 
We can understand why (\ref{eq:conventional_flow}) is called the gradient flow as follows: Let us pick up the standard (K\"ahler) metric $\diff s^2=\delta_{i\overline{j}}\diff z^i\otimes \diff \overline{z^j}$, then (\ref{eq:conventional_flow}) is indeed the gradient flow with the height function $h(z,\overline{z})=2\mathrm{Re}(S(z))$, 
\be
{\diff z^i\over \diff t}=\delta^{i\overline{j}}{\p\over \p \overline{z^j}}h(z,\overline{z}). 
\ee 
Therefore, $\mathrm{Re}(S(z))={1\over 2}h(z,\overline{z})$ monotonically increases along the flow (\ref{eq:conventional_flow}). Another important property of (\ref{eq:conventional_flow}) is that $\mathrm{Im}(S(z))$ is constant along the flow,
\be
{\diff \over \diff t}\mathrm{Im}(S(z))=0,
\ee
because of the holomorphy of $S(z)$. 

Let $\{z_{\sigma}\}_{\sigma\in\Sigma}$ be the set of the saddle points, $\p S(z_{\sigma})=0$. Using the gradient flow (\ref{eq:conventional_flow}), we define the Lefschetz thimble and its dual by~\cite{Witten:2010cx, Witten:2010zr,Cristoforetti:2012su,Fujii:2013sra, Tanizaki:2014xba} 
\be
\mathcal{J}_{\sigma}=\{z(0)\in\mathbb{C}^n\,|\, z(-\infty)=z_{\sigma}\},\, 
\mathcal{K}_{\sigma}=\{z(0)\in\mathbb{C}^n\,|\, z(+\infty)=z_{\sigma}\},
\ee
respectively. The claim from the Picard--Lefschetz theory is that we can compute relative homologies as~\cite{pham1983vanishing, Kaminski1994, Howls2271,Witten:2010cx, Witten:2010zr}   
\bea
H_n(\mathbb{C}^n,\{\mathrm{e}^{-\mathrm{Re}(S)}\ll 1\})&\simeq& \sum_{\sigma}\mathbb{Z}\, \mathcal{J}_{\sigma}, \\
H_n(\mathbb{C}^n,\{\mathrm{e}^{-\mathrm{Re}(S)}\gg 1\})&\simeq& \sum_{\sigma}\mathbb{Z}\, \mathcal{K}_{\sigma},
\eea
if $S(z)$ satisfy certain properties. By imposing appropriate orientations to $\mathcal{J}_{\sigma}$ and $\mathcal{K}_{\sigma}$, the intersection pairing satisfies 
\be
\langle \mathcal{J}_{\sigma},\mathcal{K}_{\tau}\rangle =\delta_{\sigma\tau}, 
\ee
and thus we can compute the homology class of the original integration cycle $\mathbb{R}^n$ as
\be
[\mathbb{R}^n]=\sum_{\sigma\in\Sigma}\langle \mathbb{R}^n,\mathcal{K}_{\sigma}\rangle [\mathcal{J}_{\sigma}], 
\ee
where $[\mathcal{J}]$ represents the homology class of the cycle $\mathcal{J}$.
As a result, we can rewrite the original integration as 
\be
Z=\sum_{\sigma}\langle\mathbb{R}^n,\mathcal{K}_{\sigma}\rangle\int_{\mathcal{J}_{\sigma}}\diff^n z \exp\left[-S(z)\right]. 
\label{eq:thimble_decomp}
\ee
Since $\mathrm{Im}(S(z))$ is constant along each Lefschetz thimble $\mathcal{J}_{\sigma}$, the sign problem of (\ref{eq:thimble_decomp}) can be absent or much milder than that of the original integral (\ref{eq:path_integral}). 

There is a practical way to realize the decomposition (\ref{eq:thimble_decomp}), and we introduce it following Refs.~\cite{Alexandru:2015sua, Alexandru:2016gsd}. 
Let $z(t,x)$ be the solution of (\ref{eq:conventional_flow}) with the initial condition $z(0,x)=x$. We fix the flow time $T$, and define the $n$-dimensional submanifold by 
\be
\mathcal{J}(T)=\{z(T,x)\in\mathbb{C}^n\,|\, x\in\mathbb{R}^n\}. 
\label{eq:practical_multi_thimbles}
\ee
Thanks to Cauchy's theorem, we obtain 
\bea
Z=\int_{\mathcal{J}(T)}\diff z \exp[-S(z)]=\int_{\mathbb{R}^n}\diff^n x\, \mathrm{det}\left({\p z^i(T,x)\over \p x^j}\right)\exp[-S(z(T,x))]. 
\label{eq:generalized_thimble}
\eea
The first identity means that $\mathcal{J}(T)$ belongs to the same homology class as $\mathbb{R}^n$. 
Furthermore, if $T$ is sufficiently large, $\mathcal{J}(T)$ would become almost identical to the sum of Lefschetz thimbles. Therefore, the last expression of (\ref{eq:generalized_thimble}) can be regarded as a realization of the Lefschetz-thimble decomposition (\ref{eq:thimble_decomp}) when $T$ is large enough, which is useful for numerical computations. 

\subsection{Blow-up of conventional flows}\label{sec:blowup_example}
In order to construct $\mathcal{J}(T)$, we need to solve the gradient flow (\ref{eq:conventional_flow}) numerically accurately, and thus it is quite important to understand its properties. 
Here, we would like to point out that the blow-up of solutions is  a quite generic phenomenon for nonlinear differential equations. 
To be specific, let us consider the asymptotic behavior of the gradient flow (\ref{eq:conventional_flow}) in simple examples, and we will show that the solutions of  (\ref{eq:conventional_flow}) blow up. 

The first example is a quartic potential $S(x)=x^4$. One can regard this as a
prototype  of the scalar $\phi^4$ field theory when the fields are quite large and the mass term is negligible. 
The gradient flow (\ref{eq:conventional_flow}) is 
\be
   {\diff x\over \diff t}=4x^3,
   \label{2-13}
\ee
where we consider the case $x$ is real. We can solve this equation with the initial condition $x(0)=x_0>0$ as 
\be
x(t,x_0)={1\over \sqrt{x_0^{-2}-8t}}. 
\ee
We can readily see that $x(t)\to \infty$ as $t\nearrow {1\over 8x_0^2}$, and the solution blows up within a finite time even for this simple example. 
One must be careful with the treatment of blow-up when we apply the conventional gradient flow to construct Lefschetz thimbles numerically. 
Let us make it clear that this is quite a generic phenomenon. For that purpose, we set $k=\mathrm{deg}(S)$, then the flow equation for $r^2={\sum_i |z^i|^2}(\to \infty)$ behaves as 
\be
{\diff r\over \diff t}\sim c r^{k-1},
\ee
with some positive coefficient $c$. The qualitative behavior is hence given by $r\sim (t_c-t)^{-1/(k-2)}$ for some blow-up time $t_c$. The only exception is the case when $k=2$; the blow-up does not occur only if $S$ is Gaussian. 

In order to avoid confusion, we  emphasize that the blow-up does \textit{not} violate the identity (\ref{eq:generalized_thimble}) if the equations
are  interpreted appropriately. Returning to the example Eq. \eref{2-13},
we now fix the flow time $T$, and regard $x(T,x_0)$ as a function of the initial condition $x_0$ in the following way;
\be
x(T,x_0)=\left\{
\begin{array}{ccc}
+\infty,&& x_0>1/\sqrt{8T},\\
x_0/\sqrt{1-8Tx_0^2},&& -1/\sqrt{8T}< x_0<1/\sqrt{8T},\\
-\infty, && x_0<-1/\sqrt{8T}. 
\end{array}\right.
\ee
In this example, the formula (\ref{eq:generalized_thimble}) must be interpreted as 
\be
Z=\int_{-1/\sqrt{8T}}^{1/\sqrt{8T}}\diff x_0 {\p x(T,x_0)\over \p x_0}\exp\left[-x(T,x_0)^4\right], 
\ee
which is true since it is obtained by change of variables. In this sense, (\ref{eq:generalized_thimble}) gives the correct answer. 

Let us give a heuristic argument why the formula holds true even with the blow-up. As we have seen, $\mathrm{Re}(S(z))$  increases monotonically
as the flow time becomes larger. Therefore, as the solution blows up, $\mathrm{Re}(S(z))$ diverges to $+\infty$, which implies that
\be
\mathrm{e}^{-S(z)}\to 0. \label{eq:action_blowup}
\ee
The region where the solution blows up within the flow time $T$ does not contribute because the Boltzmann weight vanishes.\footnote{Strictly speaking, the discussion given here is slightly imprecise. When the flow blows up, the Jacobian factor $\mathrm{det}(\p z(T,x)/\p x)$ diverges. Therefore, the suppression (\ref{eq:action_blowup}) must be strong enough to ensure that $\mathrm{det}(\p z(T,x)/\p x)\mathrm{e}^{-S}\to 0$. Here we just point out that this is the case for $S(x)=x^4$. }  
The rest in the original integration cycle, $(-1/\sqrt{8T},1/\sqrt{8T})$, covers the whole $\mathcal{J}(T)\subset\mathbb{C}^n$, which gives the same value as the original integral.
Therefore, the blow-up is not the problem of the formulation but requires a
correct treatment in numerical computations. 

Next, let us consider the following example, 
\be
Z=\int_{\mathbb{R}} \diff x (1-x^2)\exp\left(-x^2\right), 
\ee
i.e., $S(x)=x^2-\ln(1-x^2)$. Here, one can think of factor $(1-x^2)$ as a toy
 fermion determinant. We will apply Lefschetz-thimble method to this case, and find that Lefschetz thimbles terminate not only at infinities of the configuration space but also at the  zeros of the fermion determinant~\cite{Tanizaki:2014tua, Kanazawa:2014qma}. 
In this example, there are no infinities, so let us consider the behavior near zeros of the fermion determinant and set $x=1+\delta x$ and $|\delta x|\ll 1$. The conventional flow equation for  $\delta x$ is 
\be
{\diff \delta x\over \diff t}=-{1\over \delta x}+\cdots. 
\ee
Here, the ellipsis represents the nonsingular terms at $\delta x=0$, and we neglect them. The solution with the initial condition $\delta x(0)=\delta x_0\ll 1$ is given by 
\be
\delta x(t)=\delta x_0\sqrt{1-{2t\over \delta x_0^2}}, 
\ee
and the flow again reaches the singular point $\delta x=0$ (or $x=1$) within the finite time $t={1\over 2}\delta x_0^2$. 
To interpret the formula (\ref{eq:generalized_thimble}) correctly, at flow time $T$ we must
exclude  the region $[-1-\sqrt{2T},-1+\sqrt{2T}]\cup[1-\sqrt{2T},1+\sqrt{2T}]$ from the integral by noticing that $\mathrm{Re}(S)=+\infty$ in this region.\footnote{In the case of the bosonic action, we need require that $\mathrm{det}(\p z(T,x)/\p x)\mathrm{e}^{-S}\to 0$. In the fermionic case, this requirement is too strong and not necessarily satisfied. It is enough to require that $\mathrm{det}(\p z(T,x)/\p x)\mathrm{e}^{-S}$ is bounded around zeros of the fermion determinant because the region with blow-ups shrinks to a set of measure zero. This is
  actually the case for $S(x)=x^2-\ln(1-x^2)$ since the region $[1-\sqrt{2T},1+\sqrt{2T}]$ shrinks to the point $\{1\}$ while $\mathrm{det}(\p z(T,x)/\p x)\mathrm{e}^{-S}$ remains finite. } 

%-------------------------------------------------------------------------------%
\section{Proposal of new gradient flows without blow-up}\label{sec:new_flows}
In Sec.~\ref{sec:blowup_example}, we have seen that the blow-up of conventional gradient flows happens even for very simple examples. The formula for the Lefschetz-thimble integral is still correct even when blow-up occurs, but we must carefully control that behavior when we perform numerical computations. 
In this section, we propose a new gradient flow, 
\be
{\diff z^i\over \diff t}=\mathrm{e}^{-2\mathrm{Re}(S_B(z))/\Lambda_B}{|D|^2\over |D|^2+\Lambda_F^{-2}}\overline{\left({\p S(z)\over \p z^i}\right)},\label{eq:new_flow01}
\ee
%and 
%\be
%{\diff z^i\over \diff t}={|D(z)|^2 \over (1+\mathrm{Re}(S_B(z))^2)(1+|D(z)|^2)}\overline{\left({\p S(z)\over \p z^i}\right)}. 
%\label{eq:new_flow02}
%\ee
for some regularization parameters $\Lambda_B,\Lambda_F\ge 1$. 
We here consider the case $S(z)=S_B(z)-\ln D(z)$, where $S_B(z)$ is a polynomial that mimics the bosonic action and $D(z)$ is a polynomial that mimics the fermion determinant (i.e., $S_F=-\ln D$ is the effective action for fermions). 

\subsection{Justification of new gradient flows}\label{sec:justification}

In this section, we argue that the new gradient flow (\ref{eq:new_flow01}) also defines the Lefschetz-thimble decomposition. 
To make the argument applicable to more general cases, let us consider a regular Hermitian metric\footnote{In Refs.~\cite{Witten:2010cx, Witten:2010zr}, the K\"ahler nature of the complexified space is emphasized, but our proposal does not necessarily satisfy the K\"ahler condition. If one tries to define the intersection number between $\mathcal{J}_{\sigma}$ and $\mathcal{K}_{\tau}$ when $S$ has a continuous symmetry, the K\"ahler nature becomes important to ensure its well-definedness (see Appendix~\ref{app:complex_str} for more details). Following the recent proposal in Ref.~\cite{Alexandru:2015sua}, however, one can deform the contour appropriately without using intersection numbers, and thus the Hermitian property would be sufficient for our purpose.} $\diff s^2=g_{i\overline{j}}(z,\overline{z})\diff z^i\otimes \diff \overline{z^j}$. Especially, it should be noticed that 
\be
g_{i\overline{j}}(z,\overline{z}) v^i\overline{v^j}>0
\ee
for any $z\in\mathbb{C}^n$ and $v\in T_z\mathbb{C}^n\setminus\{0\}$. 
Using the Hermitian metric, we define the gradient flow as 
\be
{\diff z^i\over \diff t}=g^{i\overline{j}}(z,\overline{z}) \overline{\left({\p S(z)\over\p z^j}\right)}. 
\label{eq:new_flow03}
\ee
Here, $g^{i\overline{j}}$ is the inverse of metric $g_{i\overline{j}}$. 
We obtain (\ref{eq:new_flow01}) by setting 
\be
g^{ij}(z,\overline{z})=\mathrm{e}^{-2\mathrm{Re}(S_B(z))/\Lambda_B}{|D|^2\over |D|^2+\Lambda_F^{-2}}\delta^{i\overline{j}}. 
\label{eq:metric_new_flow1}
\ee
One can easily check that this metric is Hermitian on $\mathbb{C}^n\setminus\{D=0\}$. 
%We obtain (\ref{eq:new_flow01}) by setting $\displaystyle g^{i\overline{j}}=\mathrm{e}^{-2\mathrm{Re}(S)}\delta^{i\overline{j}}$, and (\ref{eq:new_flow02}) by $\displaystyle g^{i\overline{j}}={|D(z)|^2\over (1+\mathrm{Re}(S_B(z))^2)(1+|D(z)|^2)}\delta^{i\overline{j}}$. Both choices are Hermitian on $\mathbb{C}^n\setminus\{D=0\}$ since strict positivity is satisfied except at zeros of $D(z)$. 

We point out that any choice of the Hermitian metric defines an equivalent Lefschetz-thimble decomposition of the exponential integral. 
Using the flow equation (\ref{eq:new_flow03}), we obtain
\be
{\diff S(z)\over \diff t}={\p S\over \p z^i}{\diff z^i \over \diff t}=g^{i\overline{j}}{\p S\over \p z^i}{\p \overline{S}\over \p\overline{z^j}}\ge 0. 
\ee
Therefore, the two most important properties of the conventional flow equation are satisfied in the general case (\ref{eq:new_flow03}): (a) $\mathrm{Re}(S(z))$  increases monotonically and stays constant only at saddle points, $\p S(z_{\sigma})=0$. 
(b) $\mathrm{Im}(S(z))$ is constant along the gradient flow. In Appendix~\ref{app:proof_equivalence} we will give a proof that all  gradient flows define an equivalent Lefschetz thimble decomposition under certain conditions on $S(z)$. 

It would be more convincing to relate the new gradient flows (\ref{eq:new_flow01}) with the conventional gradient flow. 
We have introduced two positive parameters $\Lambda_B$ and $\Lambda_F$ in the metric of the gradient flow (\ref{eq:new_flow01}), and we can obtain the conventional flow equation by taking the limit $\Lambda_B,\Lambda_F\to \infty$. 
%To relate (\ref{eq:new_flow01}), we can consider a one-parameter family of gradient flows,
%\be
%{\diff z^i\over \diff t}=\mathrm{e}^{-2\mathrm{Re}(S_B)/\Lambda}{\mathrm{e}^{2(\Lambda-1)}|D|^2\over 1+(\mathrm{e}^{2(\Lambda-1)}-1)|D|^2}\overline{\left({\p S\over \p z^i}\right)}, 
%\label{eq:new_flow03}
%\ee
%where $\Lambda\ge 1$ is a real parameter. When $\Lambda=1$, (\ref{eq:new_flow03}) is nothing but (\ref{eq:new_flow01}), and it becomes the conventional flow (\ref{eq:conventional_flow}) in the limit $\Lambda\to \infty$. 
%Similarly, let us consider a one-parameter family, 
%\be
%{\diff z^i\over \diff t}={\mathrm{e}^{2(\Lambda-1)}|D(z)|^2 \over (1+\mathrm{Re}(S_B(z))^2/\Lambda^2)(1+\mathrm{e}^{2(\Lambda-1)}|D(z)|^2)}\overline{\left({\p S(z)\over \p z^i}\right)}. 
%\label{eq:new_flow04}
%\ee
%(\ref{eq:new_flow04}) becomes (\ref{eq:new_flow02}) at $\Lambda=1$, and it formally converges to the conventional flow equation (\ref{eq:conventional_flow}) in the limit $\Lambda\to \infty$. 
In this sense, the new and conventional gradient flows are related by a continuous deformation without violating the most important properties of the conventional flow: ${\diff \over \diff t}\mathrm{Re}(S)\ge 0$ and ${\diff \over \diff t}\mathrm{Im}(S)=0$ for any $\Lambda_B,\Lambda_F\ge 1$. 
Let us emphasize that our proposal is just a single example among the
huge set of  possibilities for the  choice of $g^{i\overline{j}}$. Other choices, such as 
\be
g^{i\overline{j}}={|D|^2\over (1+\mathrm{Re}(S_B)^2/\Lambda_B^2)(|D|^2+\Lambda_F^{-2})}\delta^{i\overline{j}},
\label{eq:new_flow02}
\ee
also satisfy all the above arguments, and we shall show that both choices work nicely to prevent  blow-up in finite time. 

\subsection{Proof of the absence of blow-up}\label{sec:absence_blowup}
 
Let us check that the new gradient flows (\ref{eq:new_flow01}) \textit{does not} blow up within a finite time. 
We assume that the action takes the form $S=S_B-\ln D$. 

We should notice that the flow diverges along the direction $\mathrm{Re}(S(z))\to +\infty$ because $\mathrm{Re}(S)$ monotonically increases. There are two possibilities to realize this divergence: 
\be
\mathrm{Re}(S_B(z))\to \infty,
\ee
or 
\be
D(z)\to 0. 
\ee
In the first case, $|z|\to \infty$ since we assumed that $S_B(z)$ is polynomial. In the second case, $z$ approaches a zero of $D$ which are
located in a bounded region. 
It is sufficient for our purpose to analyze the gradient flow
in these limiting regions.

We first consider the flow defined by (\ref{eq:new_flow01}). We can write the equation as 
\be
{\diff z^i\over \diff t}=\mathrm{e}^{-2\mathrm{Re}(S_B(z))/\Lambda_B}{|D(z)|^2\over |D|^2+\Lambda_F^{-2}} \left({\p \overline{S_B(z)}\over \p \overline{z^i}}-{1\over \overline{D(z)}}{\p \overline{D(z)}\over \p \overline{z^i}}\right).
\label{eq:flow_eq_new_explicit} 
\ee
In the limit $\mathrm{Re}(S_B(z))\to +\infty$, all  factors on the right hand side except for $\mathrm{e}^{-2\mathrm{Re}(S_B)}$ shows a polynomial dependence on $z$ or $\overline{z}$. 
Since $|z|\gg 1$ in this region, this implies
that $\diff z/\diff t$ is exponentially small, which means that $z(t)\to \infty$ only logarithmically. 
In the other limit $D(z)\to 0$, we have that $\diff z/\diff t\propto D(z)$ neglecting higher order corrections. Let $\lambda$ be a zero of $D(z)$ and write $D(z)\simeq (z-\lambda)^k$ for some $k\ge 1$, then the second term in (\ref{eq:flow_eq_new_explicit}) gives ${\diff \over \diff t}|z-\lambda|\sim -|z-\lambda|^{2k-1}$, so that $|z-\lambda|\sim t^{-1/(2k-2)}$ ($|z-\lambda|\sim \exp(-ct)$ for some $c>0$ when $k=1$). 
In both limits, it an takes infinitely long time for the flow to reach the singularities.

We next consider the gradient flow with the metric (\ref{eq:new_flow02}): 
\be
{\diff z^i\over \diff t}={|D(z)|^2 \over (1+\mathrm{Re}(S_B(z))^2/\Lambda_B^2)(|D(z)|^2+\Lambda_F^{-2})}\left({\p \overline{S_B(z)}\over \p \overline{z^i}}-{1\over \overline{D(z)}}{\p \overline{D(z)}\over \p \overline{z^i}}\right).
\ee
The same analysis holds for the limits $D(z)\to 0$, and we again obtain $|z-\lambda|\sim t^{-1/(2k-2)}$ when $z$ is close to a zero $\lambda$ of $D(z)$. 
We now consider the case $\mathrm{Re}(S_B)\to \infty$ as $|z|\to\infty$. To be specific, let $S_B(z)$ is a polynomial of order $n$, then we obtain that $\diff z/\diff t\sim 1/z^{n+1}$. Therefore, $|z(t)|\sim t^{1/(n+2)}$ and it again takes an infinitely long time for the flow to diverge. 

As a result, we have analytically shown that the blow-up within finite time can be evaded for  certain choices of the Hermitian metric $g_{i\overline{j}}$ on the complexified space $\mathbb{C}^n$, and we have constructed
two specific examples in (\ref{eq:new_flow01}) and (\ref{eq:new_flow02}). 
Let us give an intuitive explanation of why the blow-up is prevented by introduction of the metric. 
For both choices, the metric $g^{i\overline{j}}$ becomes quite small if 
\be
\mathrm{Re}(S_B)\gtrsim \Lambda_B,\, |D|\lesssim 1/\Lambda_F. 
\ee
Until the flow reaches this region, the new gradient flows show qualitatively the same behavior as the conventional one. However, once the flow reaches this region, the metric decelerates the flow sufficiently, and  blow-up does not occur at a finite time. 

\subsection{On the choice of $\Lambda_B$ and $\Lambda_F$}

For practical use of our proposal in  numerical  computations, the appropriate
choice of $\Lambda_B$ and $\Lambda_F$ is important. Let us write $g^{i\overline{j}}=g \delta^{i\overline{j}}$, then the introduction of the metric effectively changes the discretization time $\Delta t$ to $g\Delta t$. 
If one uses the fourth-order Runge-Kutta method for solving the gradient flow, the error is given by $O((g\Delta t)^4)$. 

It is quite natural from this point of view to require that $g\lesssim 1$ while solving the flow starting from real configurations. 
This puts the constraints on $\Lambda_B$ as 
\be
\Lambda_B>\left|2\min_{x\in\mathbb{R}^n}\mathrm{Re}(S_B(x))\right|.
\ee 
If we know the complex saddle points $\{z_n\}$ that have nonzero intersection numbers, then the flow reaches to the most dominant saddle points with a reasonable flow time by requiring 
\be
\Lambda_B> \left|2\min_{n}\mathrm{Re}(S_B(z_n))\right|. 
\ee
It seems that there is no constraint for the upper bound of $\Lambda_B$, so one can take a sufficiently large $\Lambda_B$ that satisfies these constraints. 

For any $\Lambda_F$, the condition $g<1$ is satisfied, and thus the lower bound is not given by this consideration. There is, however, an upper bound of $\Lambda_F$ for practical use. 
Let $z_*$ be a zero of the fermion determinant, and thus $D(z)\simeq D'(z_*)(z-z_*)$. The flow equation (\ref{eq:new_flow01}) in the vicinity of $z=z_*$ reduces to 
\be
{\diff \over \diff t}(z-z_*)\simeq -\Lambda_F^2 |D'(z_*)|^2 (z-z_*). 
\ee
The solution is given by 
\be
z-z_*\propto \exp(-\Lambda_F^2 |D'(z_*)|^2 t). 
\ee
In order to solve this exponentially fast convergence, we need to require that $\Lambda_F^2 |D'(z_*)|^2\Delta t\lesssim 1$ with the discretization time step $\Delta t$. As a result, we obtain 
\be
\Lambda_F\lesssim 1/\sqrt{|D'(z_*)|^2 \Delta t}. 
\ee
Although it is difficult to evaluate $D'(z_*)$ for realistic theories, the parametric dependence on $\Delta t$ of the upper bound of $\Lambda_F$ is obtained in this way.  
%-------------------------------------------------------------------------------%
\section{Numerical tests in simple examples}\label{sec:numerics}

In this section, we compare behavior of the conventional and new gradient flow numerically for simple examples, the Airy integral, a toy model for a fermion
determinant and a one-link U(1) model.

\subsection{Airy integral}
As a first example, we consider the Airy integration, 
\be
Z=\mathrm{Ai}(1)={1\over 2\pi}\int_{\mathbb{R}}\diff x \exp\left\{\im\left({x^3\over 3}+x\right)\right\}. 
\ee
The action of this theory is $S(z)=S_B(z)=-\im\left({z^3\over 3}+z\right)$, and it has two saddle points at $z=z_{\pm}=\pm \im$. The saddle point with non-zero intersection number is $z_{+}=\im$, and the classical action at that point is $S_B(z_{+})={2\over 3}$. 

We numerically solve the gradient flow (\ref{eq:new_flow01}). Since the fermion determinant is absent, we set $\Lambda_F\to \infty$ and write $\Lambda=\Lambda_B$: 
\be
{\diff z\over \diff t}=\exp\left(-{2\over\Lambda}\mathrm{Re}(S_B)\right)\overline{\left({\p S_B\over \p z}\right)}. 
\label{eq:new_flow_Airy}
\ee
The other flows (\ref{eq:new_flow02}) give qualitatively similar behaviors, so we do not repeat our analysis. Let $z_{\Lambda}(t,x)$ be the solution with the initial condition $z_{\Lambda}(0,x)=x$, and define the complex contour 
\be
\mathcal{J}_{\Lambda}(T_{\mathrm{flow}})=\left\{z_{\Lambda}(T_{\mathrm{flow}},x)|x\in \mathbb{R}\right\}. 
\ee

\begin{figure}\centering
\includegraphics[scale=0.4]{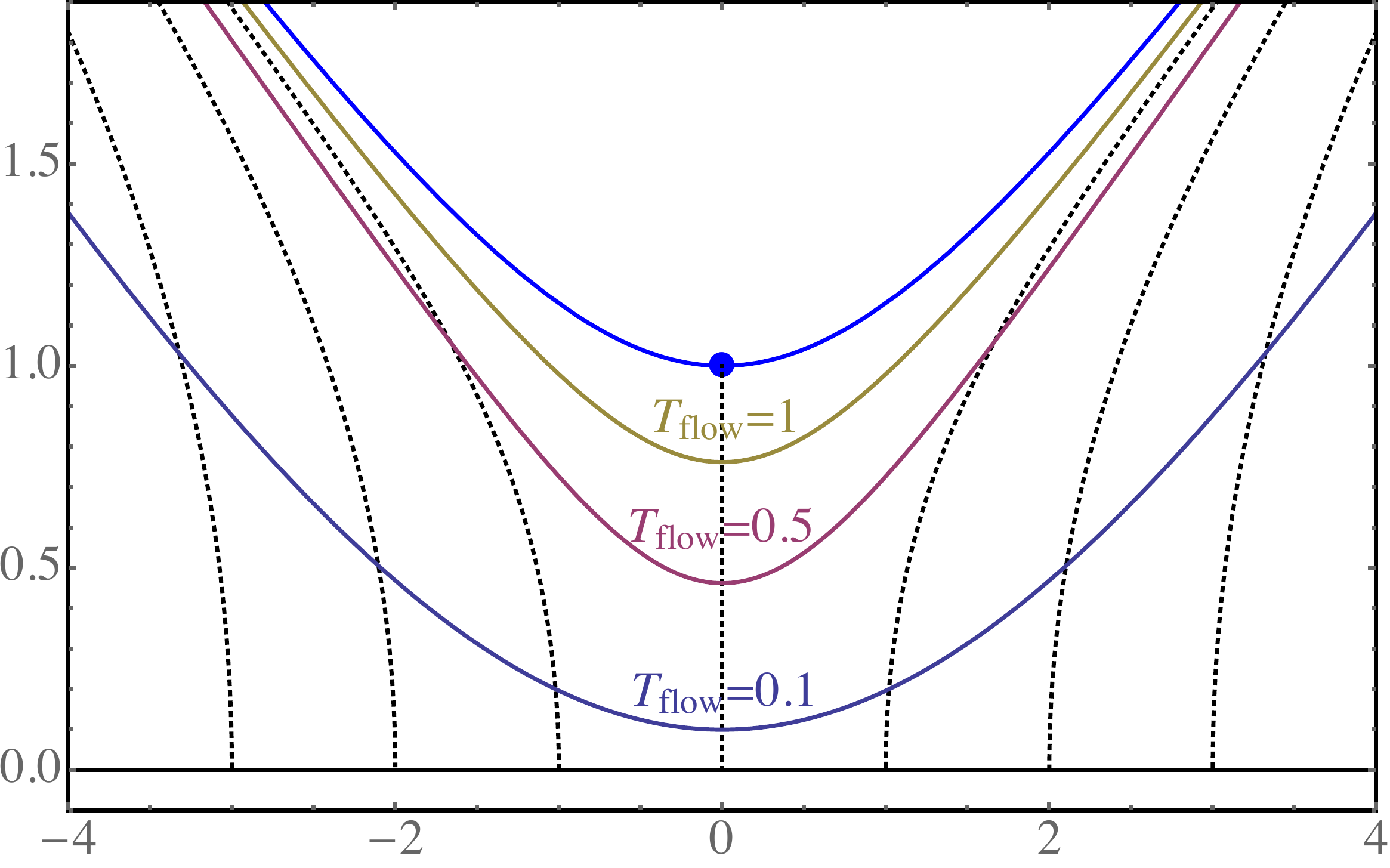}
\put(-140,-10){$\mathrm{Re}(z)$}
\put(-285,80){\rotatebox{90}{$\mathrm{Im}(z)$}}

\caption{Complex contours $\mathcal{J}_{\Lambda=\infty}(T_{\mathrm{flow}})$ for the Airy integral with the trivial metric. The dashed black curves show flow lines
with initial condition at the intersection with the real axis.}
\label{fig:Airy_flow_configs_wo_Lambda}
\end{figure}

\begin{figure}[t]\centering
\begin{minipage}{.47\textwidth}
\subfloat[$T_{\mathrm{flow}}$ dependence at $\Lambda=5$]{
\includegraphics[scale=0.29]{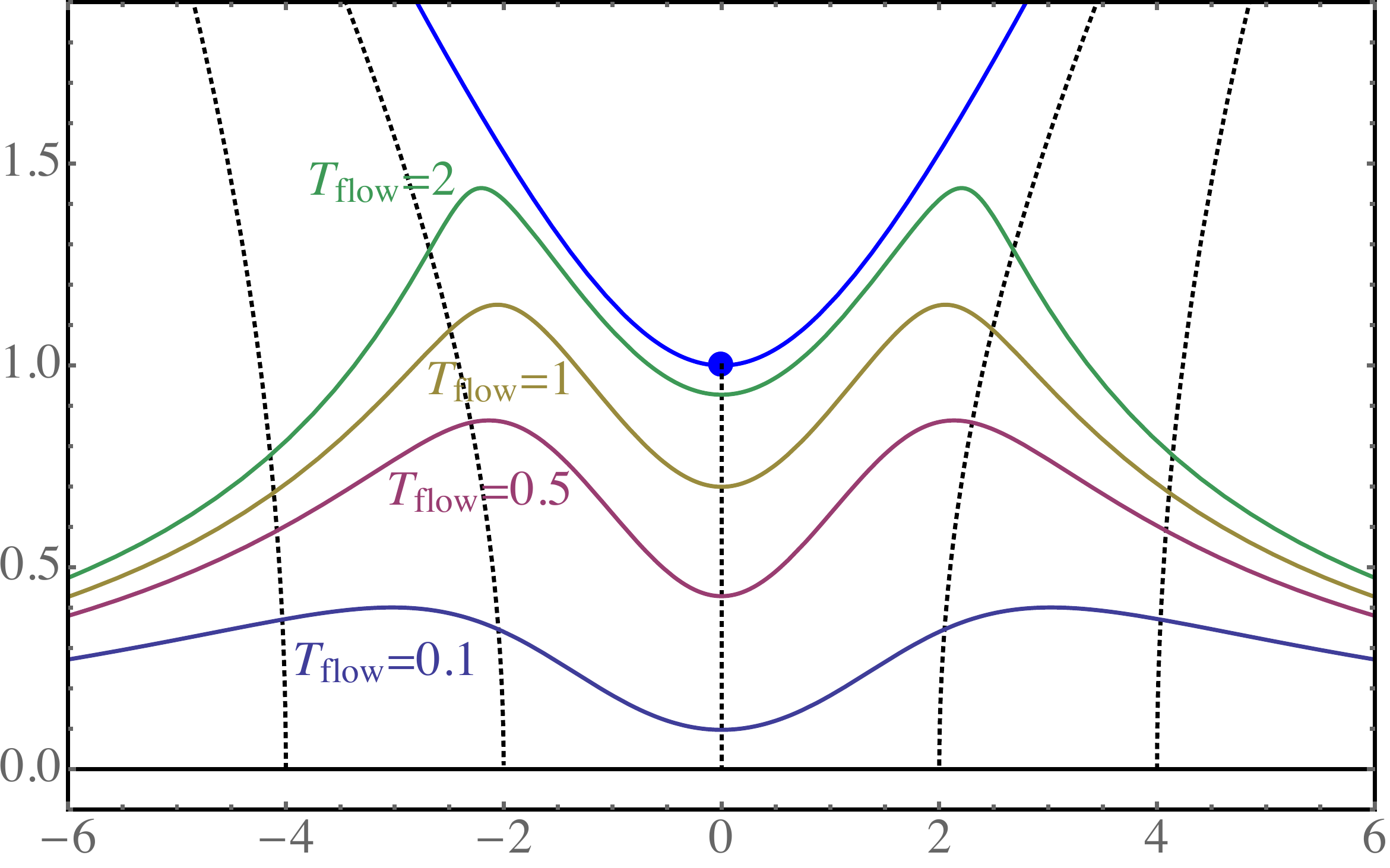}
\put(-110,-10){$\mathrm{Re}(z)$}
\put(-208,60){\rotatebox{90}{$\mathrm{Im}(z)$}}
\label{fig:Airy_flow_configs_Tdep}
}\end{minipage}\; \;
\begin{minipage}{.47\textwidth}
\subfloat[$\Lambda$ dependence at $T_{\mathrm{flow}}=1.0$]{
\includegraphics[scale=0.29]{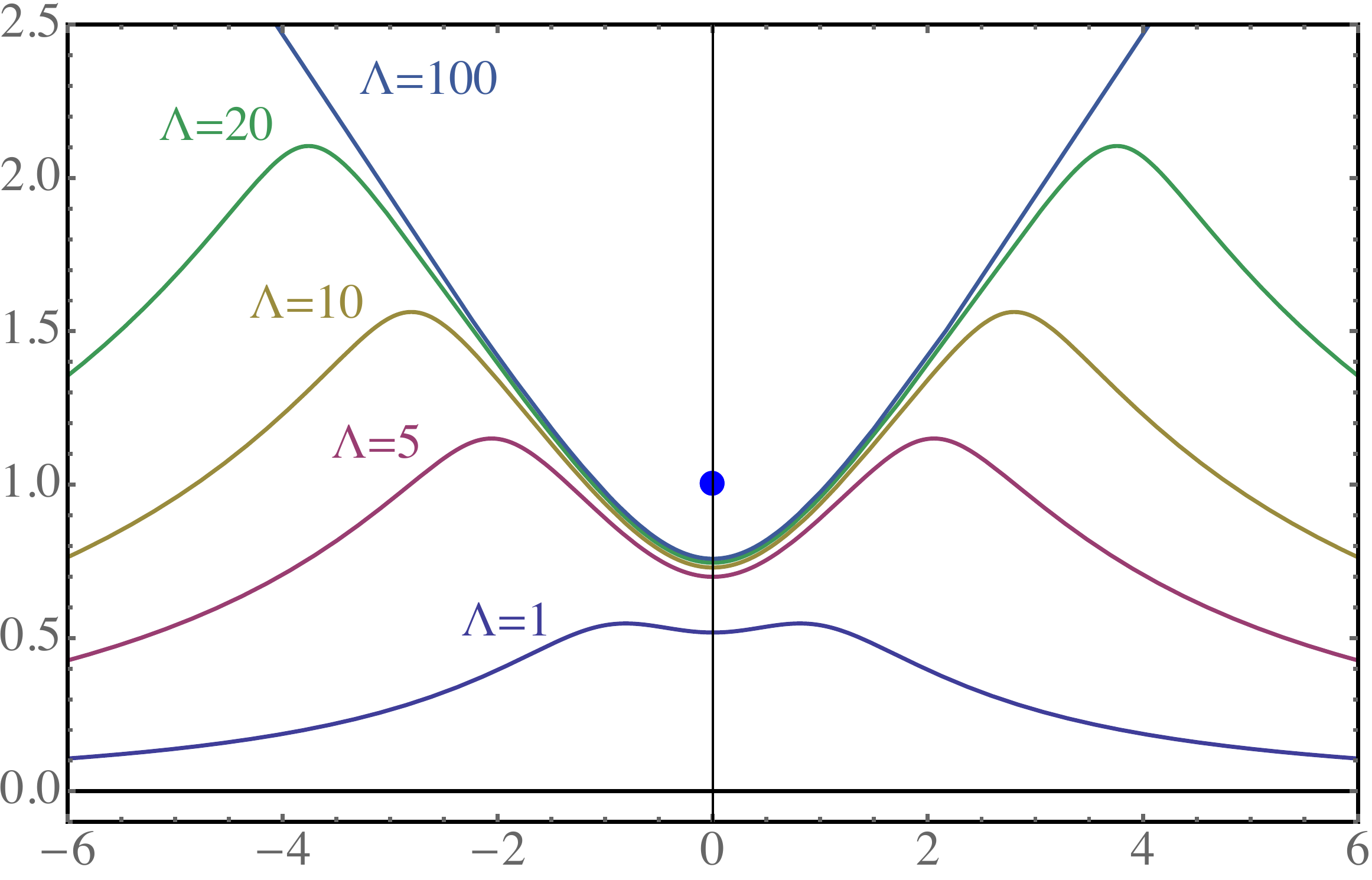}
\put(-110,-10){$\mathrm{Re}(z)$}
\put(-208,60){\rotatebox{90}{$\mathrm{Im}(z)$}}
\label{fig:Airy_flow_configs_Ldep}
}\end{minipage}
\caption{Complex contours $\mathcal{J}_{\Lambda}(T_{\mathrm{flow}})$ for the Airy integral. The black dashed curves in the left figure show flow lines with initial condition at
the intersection with the real axis. The saddle point is denoted by the blue dot which is intersected by the Lefschetz thimble in the left figure.}
\label{fig:Airy_flow_config}
\end{figure}

Figure~\ref{fig:Airy_flow_configs_wo_Lambda} shows the behavior of the gradient flow using the trivial metric, i.e. $\Lambda=\infty$. 
The solid blue line is the Lefschetz thimble. 
The dashed lines show the flow lines starting from some real points, and other solid lines show the contours $\mathcal{J}_{\Lambda=\infty}(T_{\mathrm{flow}})$ at $T_{\mathrm{flow}}=0.1$, $0.5$, and $1.0$. 
In Fig.~\ref{fig:Airy_flow_config}, we compute $\mathcal{J}_{\Lambda}(T_{\mathrm{flow}})$ at several $T_{\mathrm{flow}}$ and $\Lambda$ using fourth-order Runge-Kutta method with time step $\Delta t=0.001$. 
We show the $T_{\mathrm{flow}}$-dependence of $\mathcal{J}_{\Lambda}(T_{\mathrm{flow}})$ at $\Lambda=5$ in Fig.~\ref{fig:Airy_flow_configs_Tdep} by computing at $T_{\mathrm{flow}}=0.1$,~$0.5$,~$1.0$, and $2.0$. 
In Fig.~\ref{fig:Airy_flow_configs_Ldep}, we show the $\Lambda$-dependence at $T_\mathrm{flow}=1.0$. 
As we have shown in Sec.~\ref{sec:new_flows}, the flow equation for any $\Lambda$ prevents the blow-up since the flow slows down if $\mathrm{Re}(S_B)\gtrsim \Lambda$. 
Since $S_B(z_{+})$ is of the order of $1$, we observe in Fig.~\ref{fig:Airy_flow_configs_Ldep} that $\Lambda=1$ regularizes the flow too much and the flow stops before reaching the relevant saddle point. 
For $\Lambda\gtrsim 5$, $\mathcal{J}_{\Lambda}(T_{\mathrm{flow}}=1.0)$ almost reaches the saddle point $z_{+}$. 

\begin{figure}[t]
\centering
\includegraphics[scale=0.4]{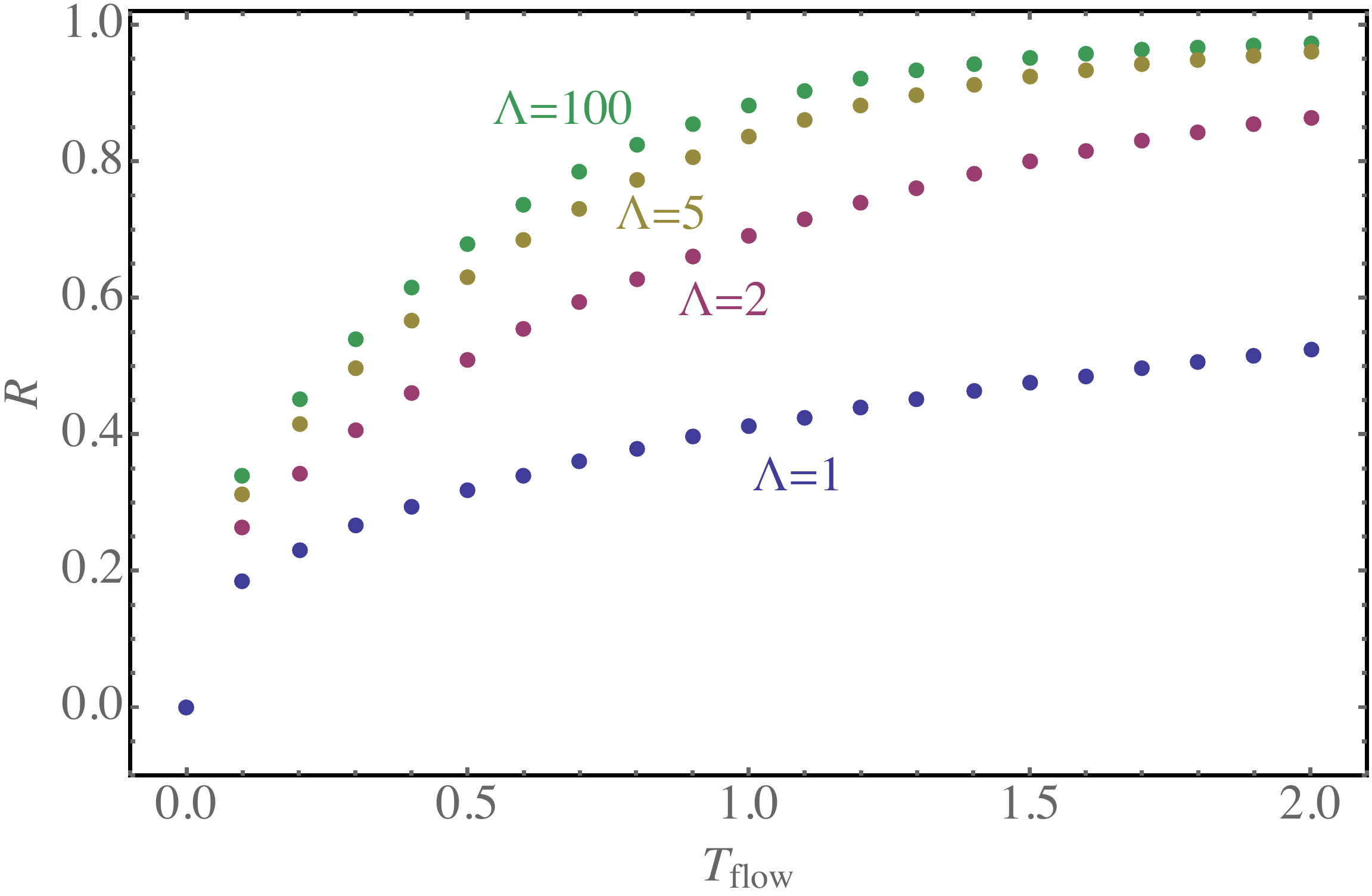}
\caption{$T_{\mathrm{flow}}$-dependence of the reweighting factor for the Airy integral at various $\Lambda$'s. }
\label{fig:Airy_reweighting}
\end{figure}

Let us now define the reweighting factor for  phase quenching on $\mathcal{J}_{\Lambda}(T_{\mathrm{flow}})$ by 
\be
R={\int_{\mathcal{J}_{\Lambda}(T_{\mathrm{flow}})}\diff z \exp(-S(z))\over \int_{\mathcal{J}_{\Lambda}(T_{\mathrm{flow}})}\left|\diff z \exp(-S(z))\right|}
={\int_{\mathbb{R}}\diff x\, {\p_x z_{\Lambda}(T_{\mathrm{flow}},x)}\exp\{-S(z(T_{\mathrm{flow}},x))\} \over \int_{\mathbb{R}}\diff x \left|{\p_x z_{\Lambda}(T_{\mathrm{flow}},x)}\exp\{-S(z(T_{\mathrm{flow}},x))\}\right|}. 
\label{eq:reweighting_factor}
\ee
When $\mathrm{T_{\mathrm{flow}}}=0$, then $\mathcal{J}_{\Lambda}(0)=\mathbb{R}$ and this reweighting factor vanishes for the Airy integral, $R(T_{\mathrm{flow}})=0$. In order to get a better understanding of the qualitative behavior of $R$, let us comment on the semiclassical evaluation of $R$ in the Lefschetz-thimble method. In this case, only the Lefschetz thimble at $z=z_{+}$ contributes. Therefore, 
\be
R(T_{\mathrm{flow}}\to\infty)\simeq {\sqrt{1/S''(z_+)}\mathrm{e}^{-S(z_+)}\over \Bigl|\sqrt{1/S''(z_+)}\mathrm{e}^{-S(z_+)}\Bigr|}=1
\ee
in the semiclassical approximation, and we expect that $R$ becomes slightly smaller in the exact computation due to the residual sign problem coming from the Jacobian factor ${\p z_{\Lambda}(T_{\mathrm{flow}},x)\over \p x}$. 
In Fig.~\ref{fig:Airy_reweighting},  we show the dependence of the reweighting factor on $T_{\mathrm{flow}}$ between $0<T_{\mathrm{flow}}\le 2.0$ at $\Lambda=1$,~$2$,~$5$, and $100$. 
In all cases, $R$ grows monotonically as $T_{\mathrm{flow}}$ becomes larger. Since the choice $\Lambda=1$ regularizes the flow too much as we have discussed, $R$ is not saturated for $T_{\mathrm{flow}}\le 2.0$. 
When $\Lambda\gtrsim 5$ for the Airy integral, the additional factor $\mathrm{e}^{-2\mathrm{Re}(S_B)/\Lambda}$ in the flow equation decelerates the flow only inside unimportant complex domains, and $R$ shows dependence on $\Lambda$ very weakly. 

\subsection{Gaussian model with fermion determinant}
We consider the following Gaussian integral:
\be
Z=\int_{\mathbb{R}}\diff x \left(x+\im (\alpha-\beta)\right)^p \exp\left(-{1\over 2}(x-\im \beta)^2\right). 
\ee
In this model, the bosonic action is $S_B(z)={1\over 2}(z-\im \beta)^2$ and the fermionic determinant is $D(z)=(z+\im (\alpha-\beta))^p$, so this model is a prototype of the sign problem with the fermion determinant. Properties of the sign problem of this model (at $\beta=0$) have been studied with the complex Langevin method in Ref.~\cite{Nishimura:2015pba}. The saddle points of the action $S=S_B-\ln D$ are given by 
\be
z=z_{\pm}=\im \beta+{-\im\alpha\pm\sqrt{4p-\alpha^2}\over 2}. 
\ee
For $\alpha<2\sqrt{p}$, which we refer to as case 1,
both the saddle points contribute to the integral, and the complex Langevin method breaks down in generic cases~\cite{Nishimura:2015pba}. This failure can be understood as a result of different complex phases for those two Lefschetz thimbles (at least within the semiclassical regime)~\cite{Hayata:2015lzj}, which necessarily requires a polynomial tail of the complex Langevin distribution and violates assumptions in the formal proof of the complex Langevin method~\cite{Aarts:2009uq,Aarts:2011ax,Nishimura:2015pba,Nagata:2016vkn} (see also Refs.~\cite{Abe:2016hpd, Salcedo:2016kyy} for recent related analytical studies). 
For $\alpha>2\sqrt{p}$, both classical solutions are purely imaginary, and only one of the saddle points has non-zero intersection number. In this case, which we refer to as case 2,
the complex Langevin method works~\cite{Nishimura:2015pba}. 
\subsubsection{Case 1}
In the following, we set $p=2$, $\alpha=2$, and $\beta=3$ so that $\alpha^2<4p$: 
\be
z_{\pm}=\pm1+2\im. 
\ee
The zero of the fermion determinant is located at $z_*=\im (\beta-\alpha)=\im$. The values of the classical action at $z=z_{\pm}$ are 
\be
S(z_{\pm})= -\ln(2)\mp \left(1+{\pi\over 2}\right)\im. 
\ee
We see from this expression that the two saddle points indeed have different phases. 

Since the Gaussian bosonic action $S_B={1\over 2}(x-\im\beta)^2$ does not cause the blow-up, we set $\Lambda_B\to\infty$ and concentrate studying the effect of the fermion determinant. We write $\Lambda=\Lambda_F$, and the flow equation becomes 
\be
{\diff z\over \diff t}={|D(z)|^2\over |D(z)|^2+\Lambda^{-2}}\left({\p \overline{S_B(z)}\over \p \overline{z^i}}-{1\over \overline{D(z)}}{\p \overline{D(z)}\over \p \overline{z^i}}\right).
\label{eq:flow_Gaussian_model}
\ee
We again solve this gradient flow for various $\Lambda$ using the fourth-order Runge-Kutta method with the time step $\Delta t=0.01$, and obtain $\mathcal{J}_{\Lambda}(T_{\mathrm{flow}})$. 

\begin{figure}[t]\centering
\begin{minipage}{.45\textwidth}
\subfloat[$T_{\mathrm{flow}}$ dependence  at $\Lambda=2$.]{
\includegraphics[scale=0.29]{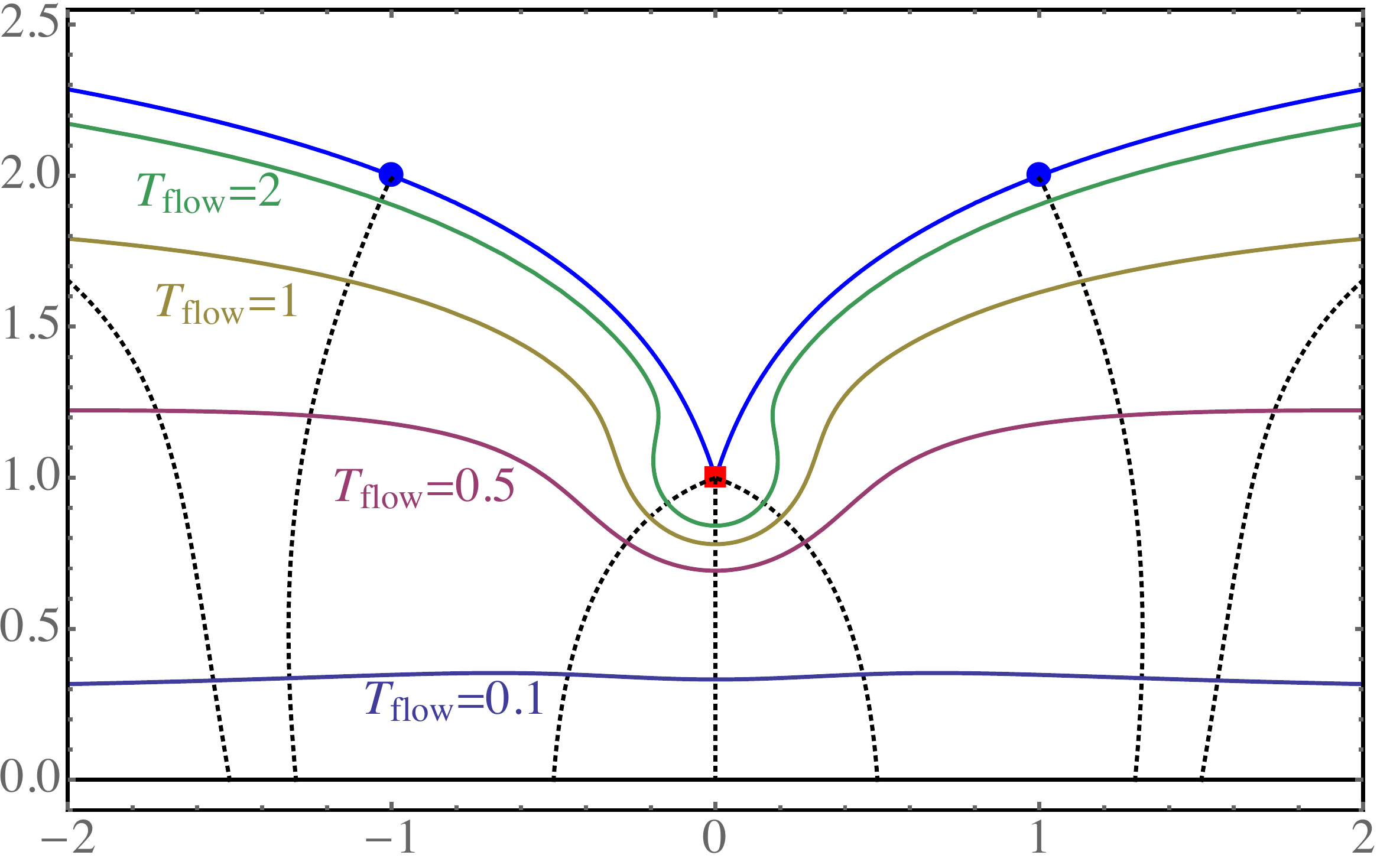}
\put(-110,-10){$\mathrm{Re}(z)$}
\put(-208,60){\rotatebox{90}{$\mathrm{Im}(z)$}}
\label{fig:Gaussian_flow_configs_Tdep}
}\end{minipage}\quad \;
\begin{minipage}{.45\textwidth}
  \subfloat[$\Lambda$ dependence of $\mathcal{J}_{\Lambda}(T_{\mathrm{flow}})$  at $T_{\mathrm{flow}}=1.0$. ]{
\includegraphics[scale=0.29]{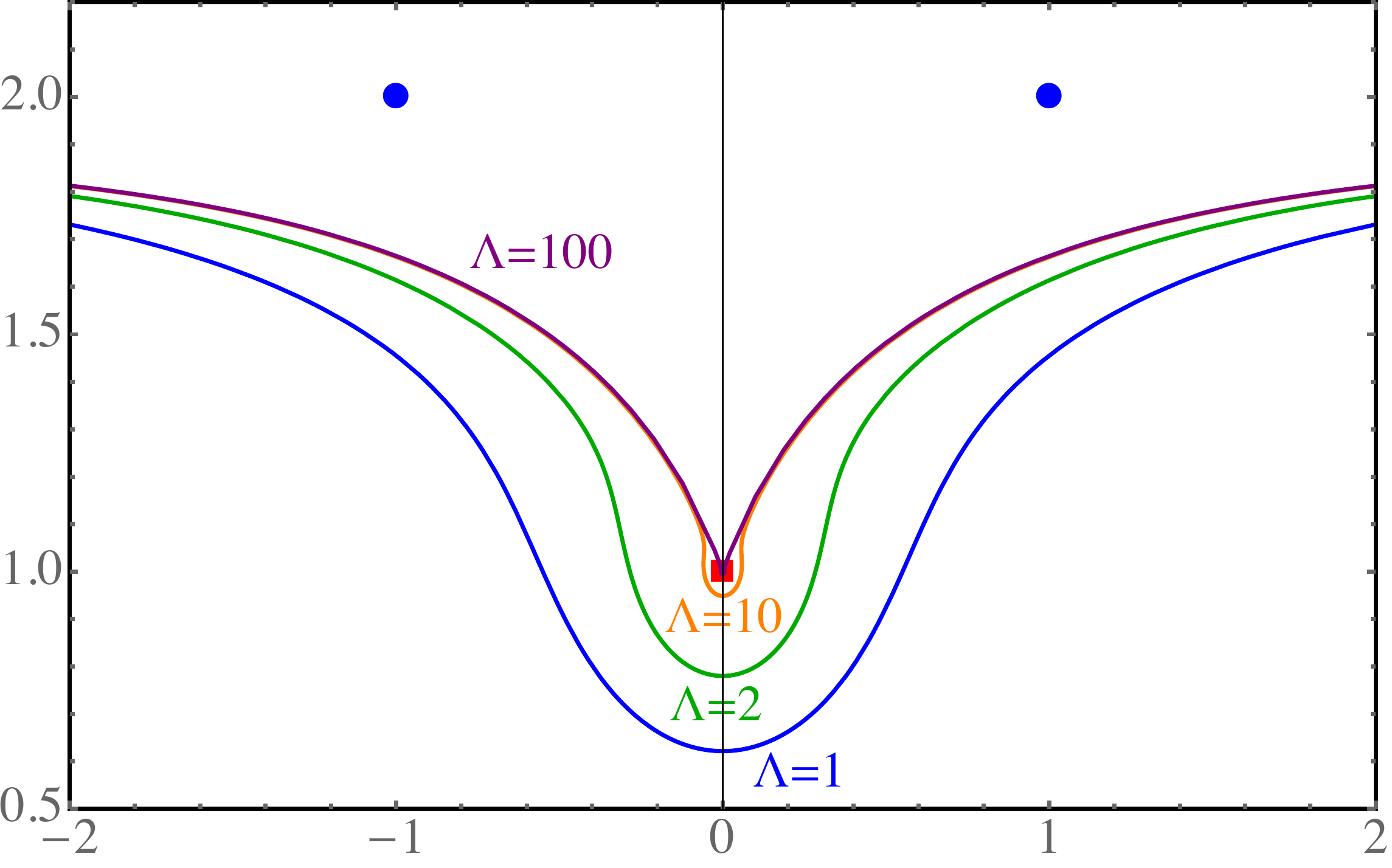}
\put(-110,-10){$\mathrm{Re}(z)$}
\put(-205,60){\rotatebox{90}{$\mathrm{Im}(z)$}}
\label{fig:Gaussian_flow_configs_Ldep}
}\end{minipage}
\caption{Complex contours $\mathcal{J}_{\Lambda}(T_{\mathrm{flow}})$ for the Gaussian model with the fermion determinant with $p=2$, $\alpha=2$, and $\beta=3$ (Case 1).
  The black dashed curves (left figure) show flow lines with initial condition at the intersection with the real axis. The zero of the ``fermion determinant'' is
  denoted by the red square while the blue dots shows the saddle points which are
intersected by the Lefschetz thimble in the left figure.}
\label{fig:Gaussian_flow_configs}
\end{figure}

In Fig.~\ref{fig:Gaussian_flow_configs}, we show how $\mathcal{J}_{\Lambda}(T_{\mathrm{flow}})$ develops as $T_{\mathrm{flow}}$ and $\Lambda$ are changed. 
In Fig.~\ref{fig:Gaussian_flow_configs_Tdep}, its $T_{\mathrm{flow}}$-dependence at $\Lambda=2$ is shown, and $\mathcal{J}_{\Lambda}(T_{\mathrm{flow}})$ approaches to the saddle point, $z_{\pm}=\pm1+2\im$ as $T_{\mathrm{flow}}$ becomes larger. 
Let us also pay attention to the behavior of flows in the vicinity of the zero of $D$, $z_*=\im$. Since the flow slows down around $z=z_*$, the complex contours $\mathcal{J}_{\Lambda}(T_{\mathrm{flow}})$ make a slight detour to evade that point. 
This feature can be more clearly seem by looking at the $\Lambda$-dependence of the contours.  In Fig.~\ref{fig:Gaussian_flow_configs_Ldep}, the $\Lambda$-dependence is studied at $T_{\mathrm{flow}}=1.0$, and the detour becomes smaller as $\Lambda$ becomes larger. 
This is consistent with the previous analysis that the flow decelerates if $|D|\lesssim \Lambda^{-1}$. 

\begin{figure}[t]
\centering
\includegraphics[scale=0.6]{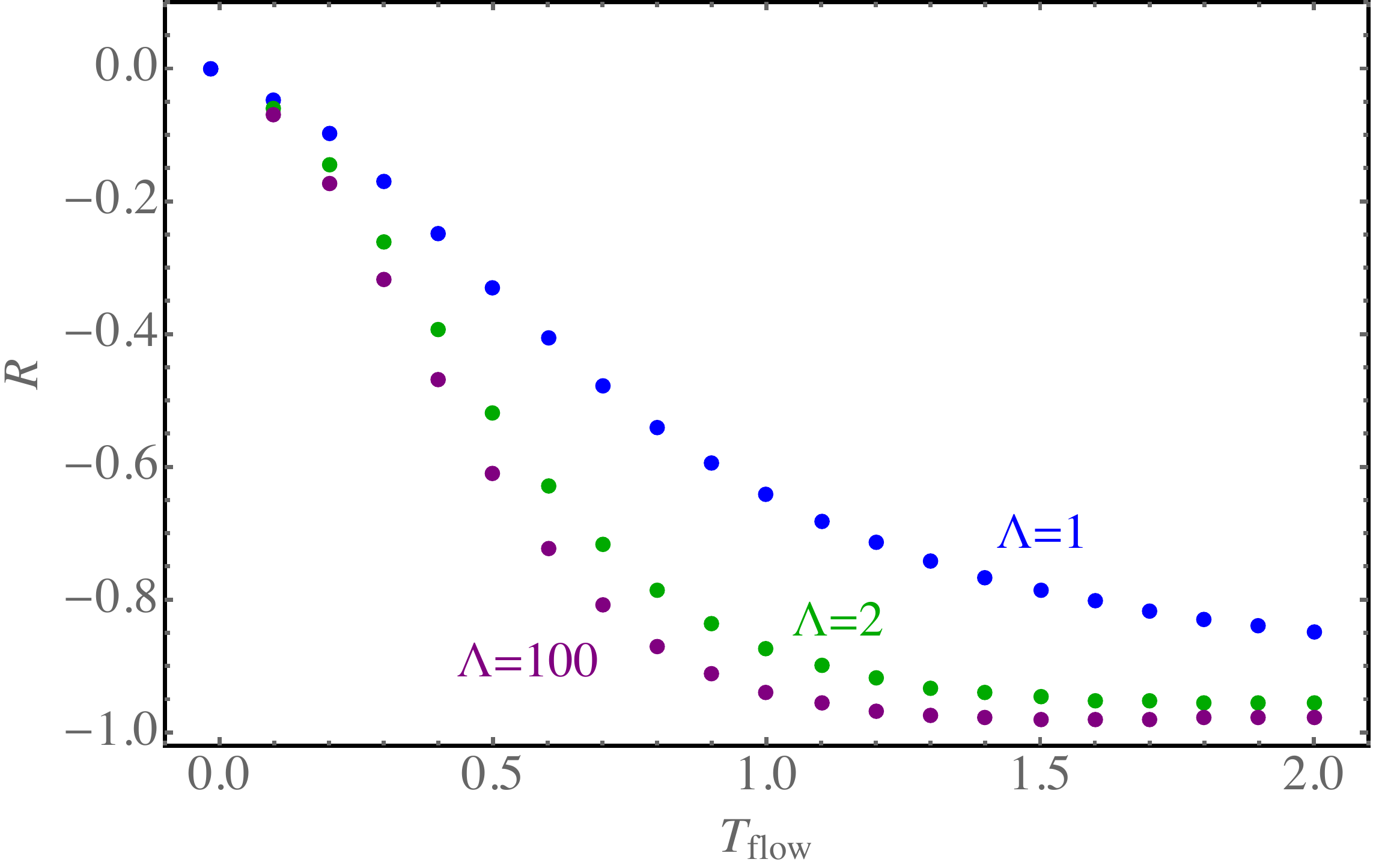}
\caption{$T_{\mathrm{flow}}$-dependence of the reweighting factor at various $\Lambda$'s for the Gaussian model for $p=2$, $\alpha=2$, and $\beta=3$ (Case 1). }
\label{fig:Gaussian_reweighting}
\end{figure}
Figure~\ref{fig:Gaussian_reweighting} shows the $T_{\mathrm{flow}}$-dependence of the reweighting factor $R$ at $\Lambda=1$, $2$, and $100$ for $0<T_{\mathrm{flow}}\le 2.0$. Since the partition function $Z$ is negative with our setup, the reweighting factor $R$ is also negative in this case. The reweighting factor of the conventional phase quenching, i.e. $R(T_{\mathrm{flow}}=0)$, is about $-1.7\times 10^{-2}$ when $p=2$, $\alpha=2$, and $\beta=3$, and we can observe how $R$ develops as $T_{\mathrm{flow}}$ becomes larger for various values of $\Lambda$. To get a better understanding of the figure, let us comment on the semiclassical evaluation of the reweighting factor in the Lefschetz-thimble method.  
In the saddle-point approximation, the reweighting factor is given by 
\be
R(T_{\mathrm{flow}}\to\infty)\simeq {\sqrt{1/S''(z_+)}\mathrm{e}^{-S(z_+)}+\sqrt{1/S''(z_-)}\mathrm{e}^{-S(z_-)}\over \Bigl|\sqrt{1/S''(z_+)}\mathrm{e}^{-S(z_+)}\Bigr|+\Bigl|\sqrt{1/S''(z_-)}\mathrm{e}^{-S(z_-)}\Bigr|}. 
\ee
It is easy to find that $\sqrt{1/S''(z_\pm)}\mathrm{e}^{-S(z_\pm)}=c\exp\left(\pm \im(1+{\pi\over 2}+{\pi\over 8})\right)$ for some $c>0$. 
As a result, we get $R(T_{\mathrm{flow}}\to\infty)=\cos(1+{\pi\over 2}+{\pi\over 8})\simeq -0.98$ in the semiclassical approximation, which is roughly consistent with the saturation value given in Fig.~\ref{fig:Gaussian_reweighting}.  

\subsubsection{Case 2}
In the following, we set $p=2$, $\alpha=3$, and $\beta=4$ so that $\alpha^2>4p$, and the saddle points are located at 
\be
z_+=3\im, \ \, z_-=2\im. 
\ee
The zero of the fermion determinant is at $z=z_*=\im$. The values of the classical action at $z=z_{\pm}$ are 
\be
S(z_{+})= -{1\over 2}-2\ln(2)-\im\pi,\ \, S(z_-)=-2-\im \pi. 
\ee
These two classical actions have the same imaginary part, and the theory is indeed on the Stokes ray; two saddle points $z_{\pm}$ are connected by the gradient flow. 
We again consider the gradient flow given by (\ref{eq:flow_Gaussian_model}). 

\begin{figure}[t]\centering
\includegraphics[scale=0.45]{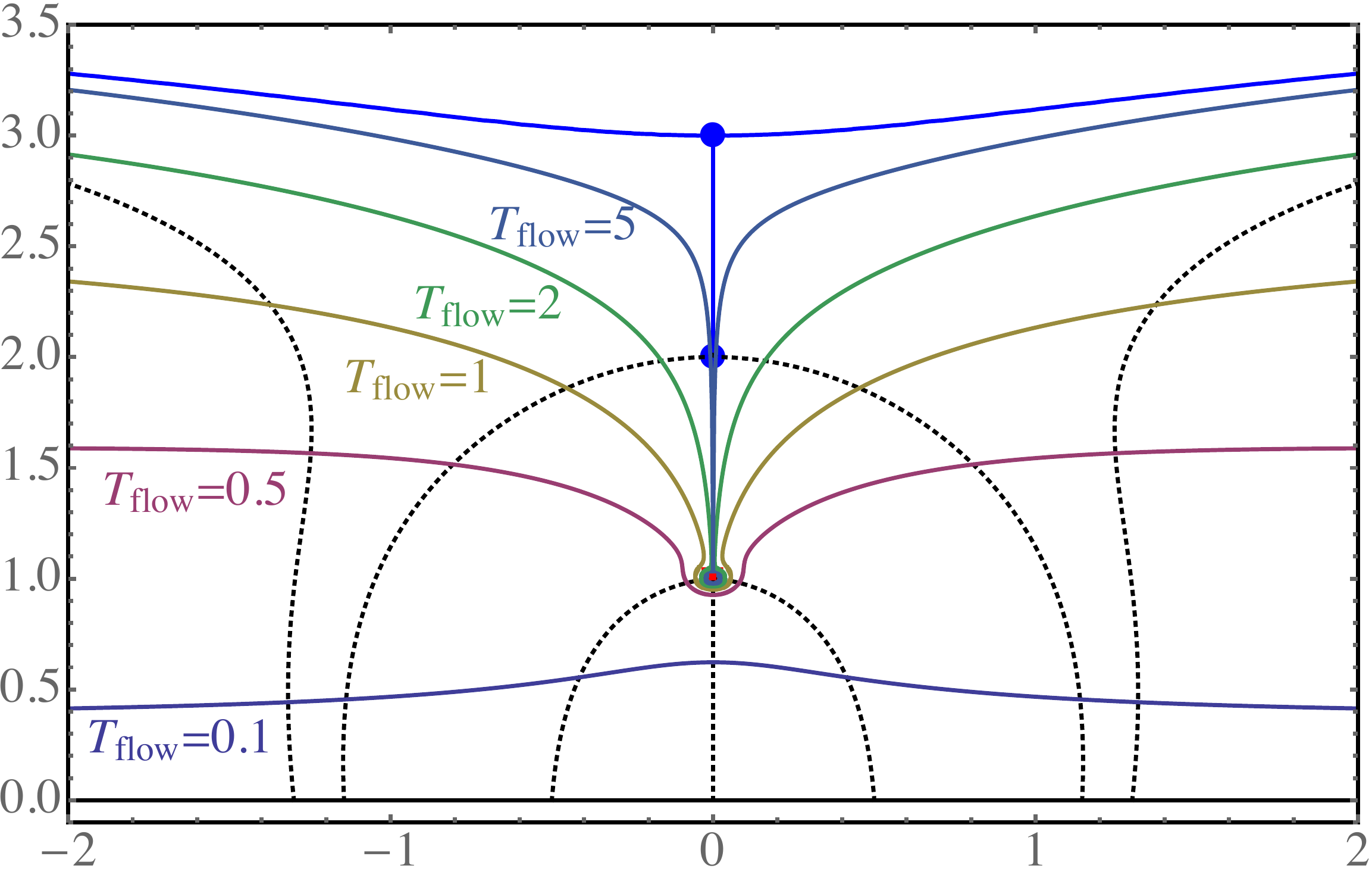}
\put(-155,-10){$\mathrm{Re}(z)$}
\put(-315,80){\rotatebox{90}{$\mathrm{Im}(z)$}}
\caption{Complex contours $\mathcal{J}_{\Lambda}(T_{\mathrm{flow}})$ for the Gaussian model with the fermion determinant with $p=2$, $\alpha=3$, and $\beta=4$ (case 2). The saddle points are again denoted by the blue dots with the saddle
  point at $3\im$ intersected by the Lefschetz thimble. The zero of the fermion
  determinant at $i$ is represented by the red square. The black dashed curves show
flow lines with initial condition at the intersection with the real axis.}
\label{fig:Gaussian_flow_configs_case2}
\end{figure}

This is a tricky example because $\mathcal{J}_{\Lambda}(T_{\mathrm{flow}})$ does not converge to the contributing Lefschetz thimble $\mathcal{J}_+$ although their homology classes are the same, $[\mathcal{J}_{\Lambda}(T_{\mathrm{flow}})]=[\mathcal{J}_+]$ ($\mathcal{J}_{\pm}$ are Lefschetz thimbles associated with saddle points $z_{\pm}$). To make the discussion on the intersection number well-defined, let us imagine that we add an infinitesimal imaginary part to parameters $\alpha$, $\beta$ so that the theory is off the Stokes ray. 
If one draws the dual thimbles $\mathcal{K}_{+}$ and $\mathcal{K}_-$, one will find that $\mathcal{K}_+$ intersects with $\mathbb{R}$ only once but that $\mathcal{K}_-$ intersects with $\mathbb{R}$ twice with different relative orientations. Their intersection number with $\mathbb{R}$ can be computed as $\langle \mathbb{R},\mathcal{K}_{+}\rangle =1$ and $\langle \mathbb{R},\mathcal{K}_{-}\rangle=1-1(=0)$. As a result, the Lefschetz-thimble decomposition of the integral becomes 
\be
\int_{\mathbb{R}}\diff x \exp(-S(x))=\int_{\mathcal{J}_++\mathcal{J}_--\mathcal{J_-}}\diff z\exp(-S(z))=\int_{\mathcal{J}_+}\diff z\exp(-S(z)). 
\ee
We can now see why $\mathcal{J}_{\Lambda}(T_{\mathrm{flow}})\not\to \mathcal{J}_+$ as $T_{\mathrm{flow}}\to \infty$ as manifolds: The construction of $\mathcal{J}_{\Lambda}(T_{\mathrm{flow}})$ is sensitive to the cancellation of two intersections between $\mathcal{K}_-$ and $\mathbb{R}$, and thus the limit of $\mathcal{J}_{\Lambda}(T_{\mathrm{flow}})$ roughly becomes 
\be
\lim_{T_{\mathrm{flow}}\to \infty}\mathcal{J}_{\Lambda}(T_{\mathrm{flow}})\simeq (-\infty+z_+,z_+-0^+]\cup[z_+-0^+,z_*]\cup[z_*,z_++0^+]\cup[z_++0^+,z_++\infty). 
\label{eq:contour_Gaussian_case2}
\ee
Since $\mathcal{J}_+\simeq (-\infty+z_+,z_++\infty)$, there are additional line segments, along which the integrals of holomorphic functions cancel.  This observation is important when we discuss the reweighting factor, since the additional segments reduce the reweighting factor: $R(T_{\mathrm{flow}}\to \infty)\simeq -0.45$. 

Since the $\Lambda$ dependence is quite weak except in the vicinity of $z=z_*=\im$ for $\Lambda\gtrsim 10$ as we have seen in Fig.~\ref{fig:Gaussian_flow_configs_Ldep} for slightly different parameters, let us show the numerical result only at $\Lambda=10$. 
Figure~\ref{fig:Gaussian_flow_configs_case2} gives the $T_{\mathrm{flow}}$ dependence of contours $\mathcal{J}_{\Lambda}(T_{\mathrm{flow}})$ at $\Lambda=10$ for $0<T_{\mathrm{flow}}\le 5.0$. 
We find that $\mathcal{J}_{\Lambda}(T_{\mathrm{flow}})$ indeed approaches the contour given in (\ref{eq:contour_Gaussian_case2}) as $T_{\mathrm{flow}}$ gets larger. 
Moreover, thanks to the metric in the gradient flow, $\mathcal{J}_{\Lambda}(T_{\mathrm{flow}})$ detours evading the zero of the fermion determinant $z=z_*$. 

\begin{figure}[t]
\centering
\includegraphics[scale=0.38]{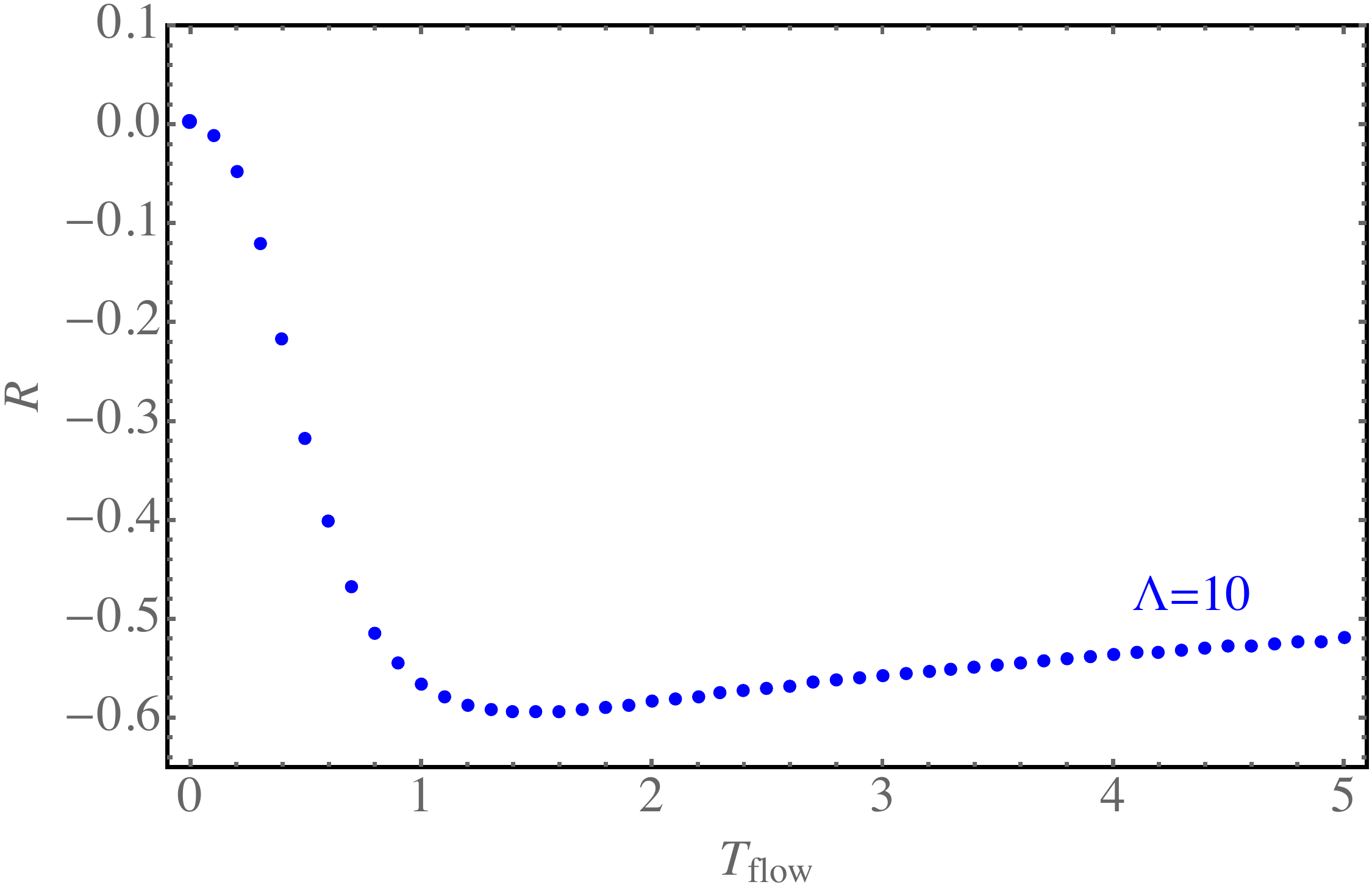}
\caption{$T_{\mathrm{flow}}$-dependence of the reweighting factor at $\Lambda=10$ for the Gaussian model for $p=2$, $\alpha=3$, and $\beta=4$. }
\label{fig:Gaussian_reweighting2}
\end{figure}

Figure~\ref{fig:Gaussian_reweighting2} shows the $T_{\mathrm{flow}}$ dependence of the reweighting factor. Interestingly, the reweighting factor reaches the maximum around $T_\mathrm{flow}\simeq 1.4$ and overcomes the reweighting factor, $R(T_{\mathrm{flow}}\to \infty)\simeq -0.45$, computed by using Lefschetz thimbles.  It gradually decreases after that, and approaches to $R(T_{\mathrm{flow}}\to \infty)\simeq -0.45$.

\subsection{$U(1)$ one-link model}
The $U(1)$ one-link model is given by 
\be
Z=\int_{-\pi}^{\pi}{\diff x\over 2\pi}\mathrm{e}^{\beta \cos(x)}[1+\kappa \cos(x-\im\mu)]. 
\label{eq:partition_function_U1}
\ee
The bosonic action is $S_B(z)=-\beta \cos(z)$, and the fermion determinant is given by $D(z)=(1+\kappa\cos(z-\im\mu))$. 
This model is considered in Ref.~\cite{Aarts:2014nxa} in the context of Lefschetz thimbles. In order to control the blow-ups of this model at various values of the parameters, we need to introduce both $\Lambda_B$ and $\Lambda_F$ in the metric of the gradient flow. We shall see that our proposal to change the flow works well also for this situation. 
The structure of Lefschetz thimbles changes drastically as the parameter $\kappa$ exceeds $1$, so we consider the cases $\kappa=1/2$ and $\kappa=2$. 
We always set $\beta=1$ and $\mu=2$. 

\subsubsection{Small $\kappa$}

\begin{figure}[t]\centering
\begin{minipage}{.48\textwidth}
\subfloat[$T_{\mathrm{flow}}$ dependence at $\Lambda_F=5$]{
\includegraphics[scale=0.29]{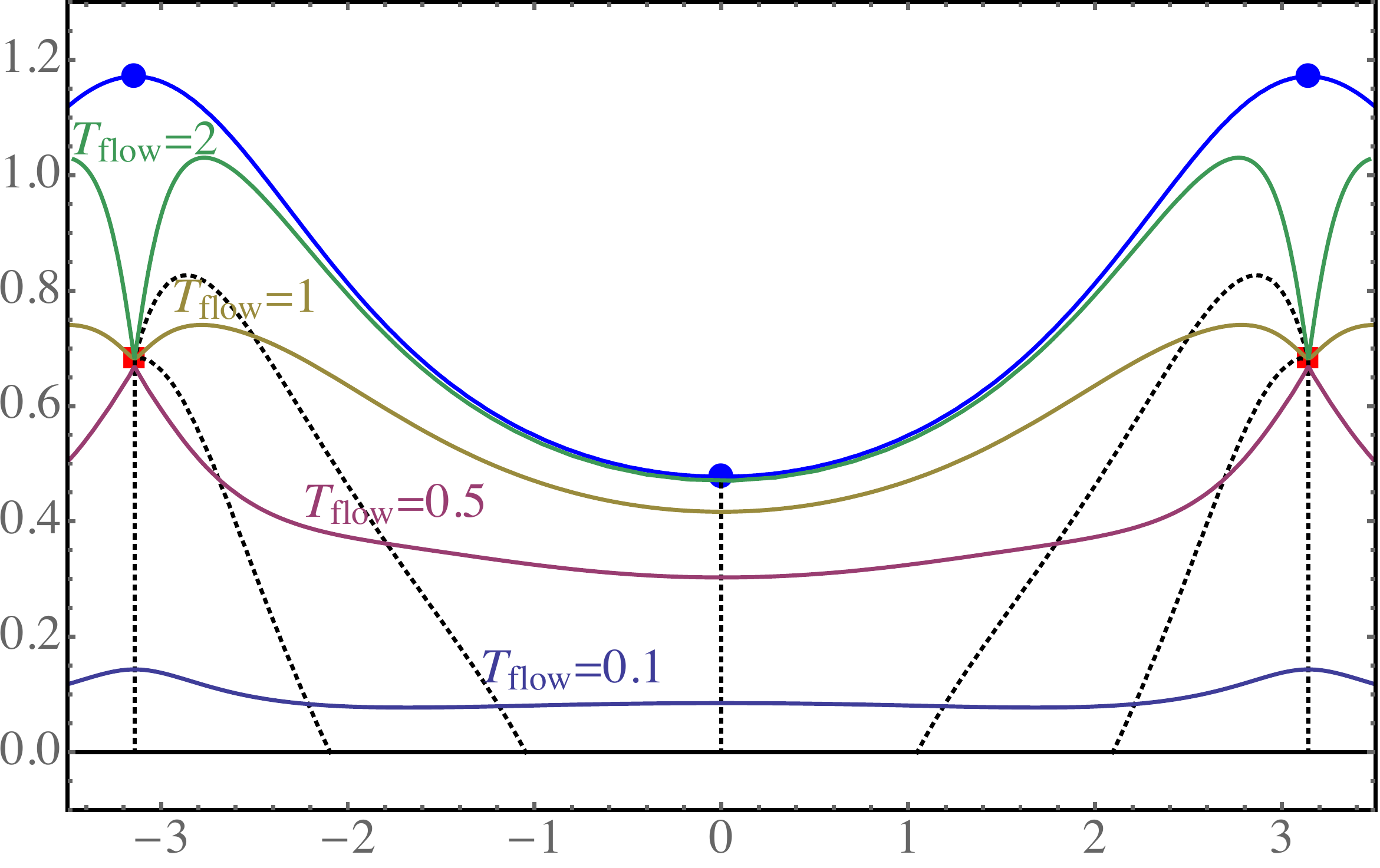}
\put(-110,-10){$\mathrm{Re}(z)$}
\put(-208,60){\rotatebox{90}{$\mathrm{Im}(z)$}}
\label{fig:U1_flow_configs_Tdep_SmallKappa}
}\end{minipage}\; \;
\begin{minipage}{.48\textwidth}
\subfloat[$\Lambda_F$ dependence at $T_{\mathrm{flow}}=1.0$]{
\includegraphics[scale=0.29]{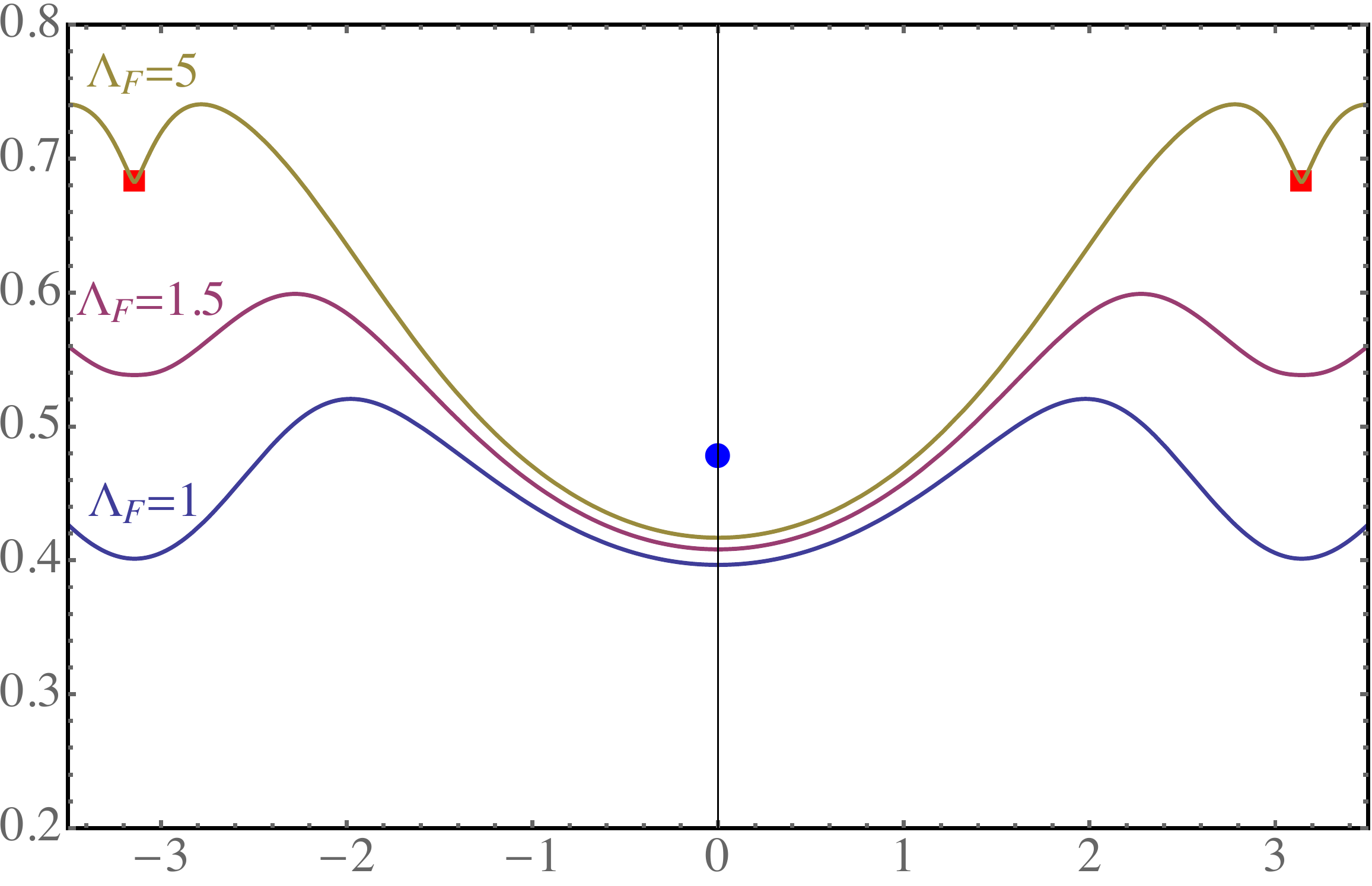}
\put(-110,-10){$\mathrm{Re}(z)$}
\put(-208,60){\rotatebox{90}{$\mathrm{Im}(z)$}}
\label{fig:U1_flow_configs_Ldep_SmallKappa}
}\end{minipage}
\caption{Complex contours $\mathcal{J}_{\Lambda}(T_{\mathrm{flow}})$ for the $U(1)$ one-link model with $\kappa=1/2$. We set $\Lambda_B=5$. The saddle points are denoted by the blue dots while the zeros of the fermion determinant are depicted as red squares. The dashed curves in the left figure are flow lines which end
  at a zero of the fermion determinant or at the saddle point on the imaginary
axis which is intersected by the Lefschetz thimble.}
\label{fig:U1_flow_config}
\end{figure}

\begin{figure}[t]
\centering
\includegraphics[scale=0.4]{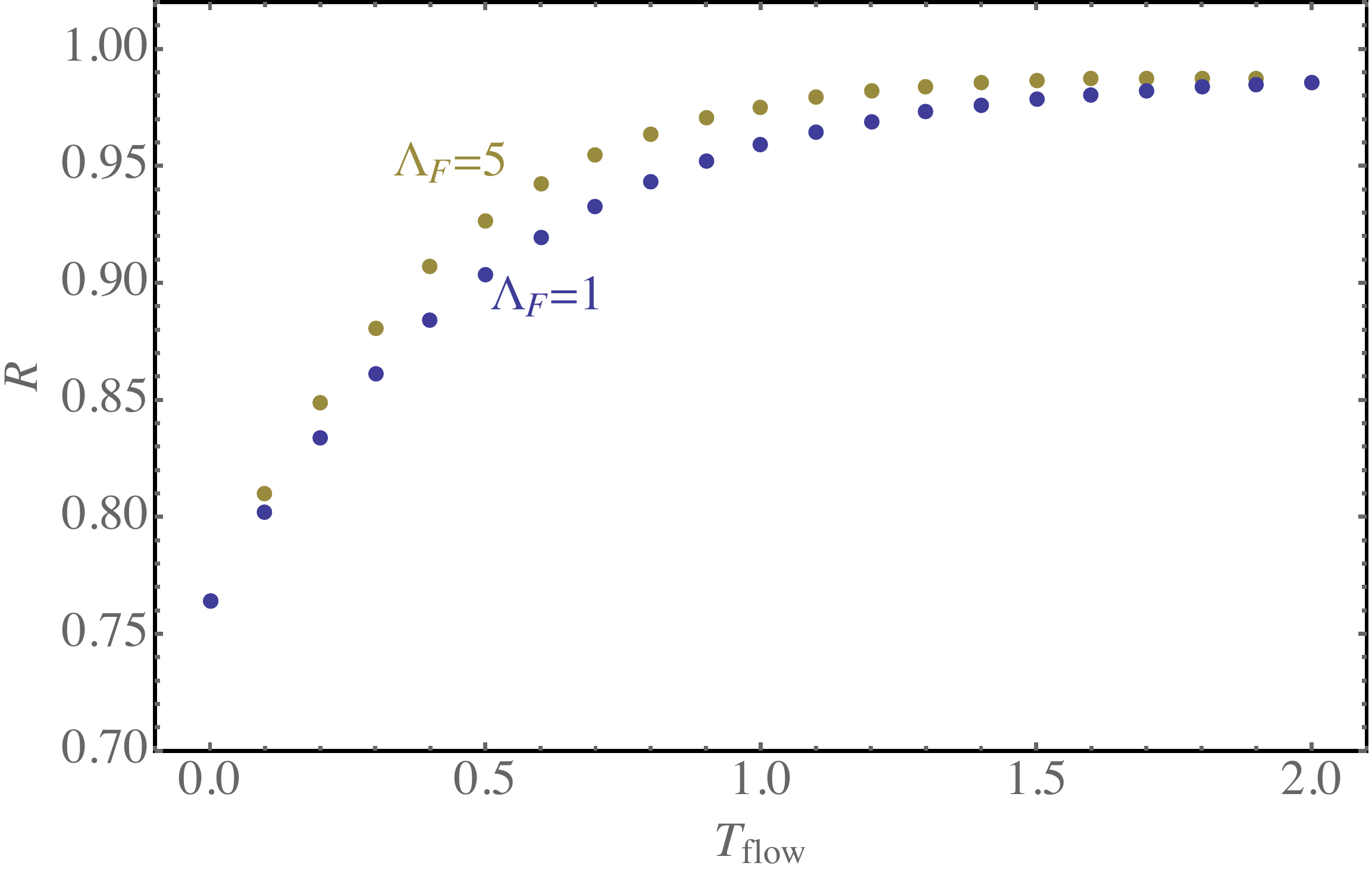}
\caption{$T_{\mathrm{flow}}$-dependence of the reweighting factor for the $U(1)$ one-link model with $\kappa=1/2$ and $\Lambda_B=5$. }
\label{fig:U1_reweighting_SmallKappa}
\end{figure}
We take $\kappa=1/2$, $\beta =1$, and $\mu =2$ in (\ref{eq:partition_function_U1}), and we set $\Lambda_B=5$ throughout the analysis in this case. The relevant saddle points are approximately given by 
\be
z_1\simeq 0.48\im,\quad z_2\simeq \pi+1.17\im. 
\ee
The values of the classical action at these saddle points are $S_1\simeq -1.9$ and $S_2\simeq 2.9$, respectively,  and thus the contribution is dominated by $z_1$. 
The zeros of the fermion determinant are located at
\be
z_{*\pm}=\pi+\im\left(\mu\pm \cosh^{-1}(1/\kappa)\right).  
\ee
In Fig.~\ref{fig:U1_flow_configs_Tdep_SmallKappa}, the blue solid curve shows the Lefschetz thimble of $z_1$ that contributes to $Z$, and red squares show the zero $z_{*-}$ of the fermion determinant. 
We show in Fig.~\ref{fig:U1_flow_configs_Tdep_SmallKappa} how $\mathcal{J}_{\Lambda}(T_{\mathrm{flow}})$ develops as $T_{\mathrm{flow}}$ increases at $\Lambda_B=\Lambda_F=5$. 
In Fig.~\ref{fig:U1_flow_configs_Ldep_SmallKappa}, the $\Lambda_F$ dependence of $\mathcal{J}_{\Lambda}(T_{\mathrm{flow}})$ is studied at $T_{\mathrm{flow}}=1.0$, and the contours approach to $z=z_{*-}$ as $\Lambda_F$ becomes larger. 
In this parameter region, $\Lambda_B$ does not play a significant role, because the blow-up due to the bosonic action $S_B$ does not occur.

Figure~\ref{fig:U1_reweighting_SmallKappa} shows the $T_{\mathrm{flow}}$-dependence of the reweighting factor at $\Lambda_F=1$ and $5$. We find that the $\Lambda_F$ dependence of the reweighting factor is quite small even though the contours themselves strongly depend of $\Lambda_F$ as we have seen in Fig.~\ref{fig:U1_flow_configs_Ldep_SmallKappa}. 
In this case, the contribution to $Z$ is dominated by one saddle point $z_1$, and thus the reweighting factor becomes close to $1$. 
For $\Lambda_F=5$, the reweighting factor reaches its maximum around $T_{\mathrm{flow}}\simeq 1.6$, and it slightly decreases after that. This is because the zero $z_{*-}$ obstructs the deformation of real cycle to the Lefschetz thimble shown by the blue solid curve as we have seen in the Gaussian model, and thus the residual sign problem becomes more severe when $T_{\mathrm{flow}}$ becomes larger than a certain value.

\subsubsection{Large $\kappa$}

\begin{figure}[t]\centering
\begin{minipage}{.48\textwidth}
\subfloat[$T_{\mathrm{flow}}$ dependence at $\Lambda_F=5$]{
\includegraphics[scale=0.29]{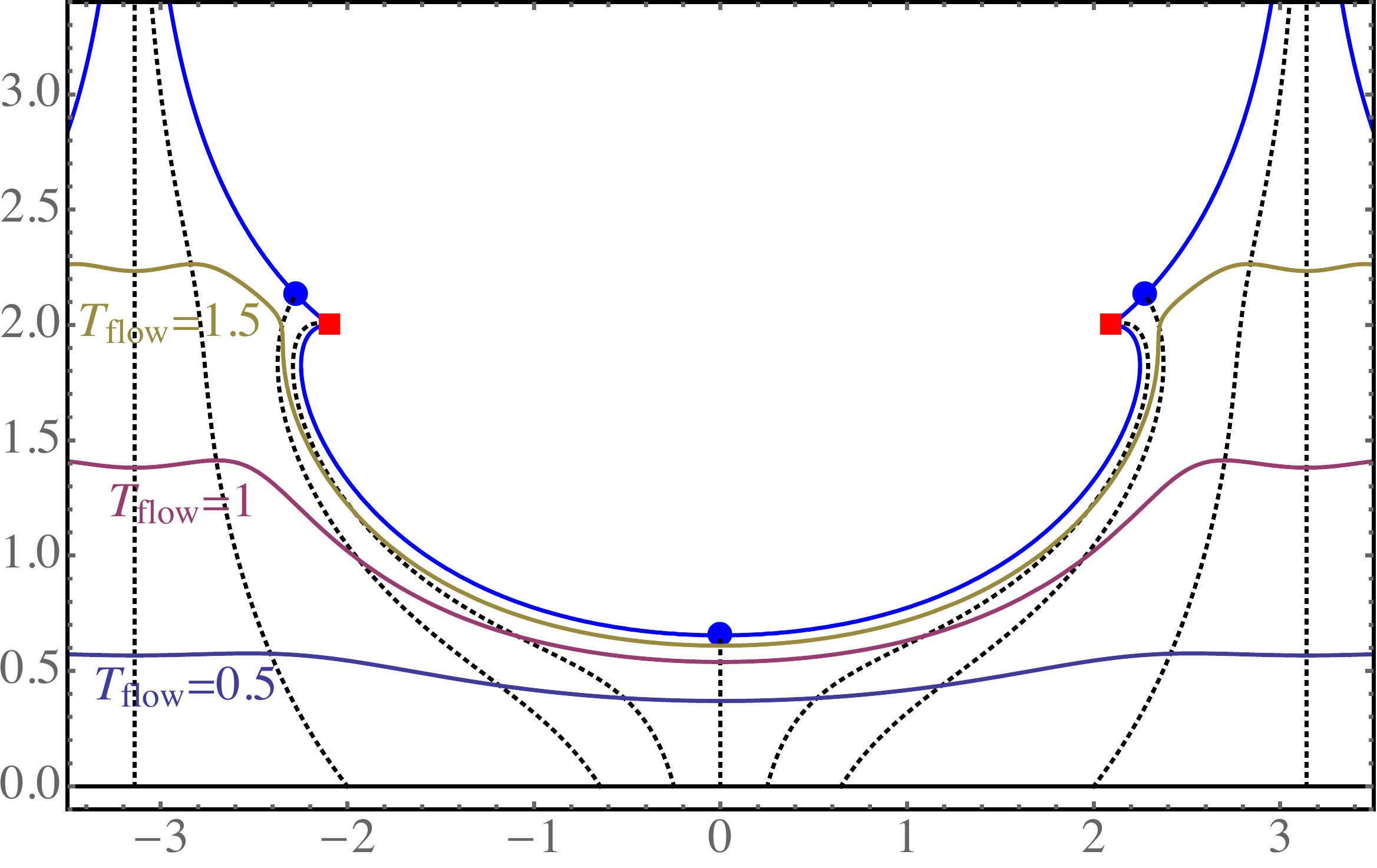}
\put(-110,-10){$\mathrm{Re}(z)$}
\put(-208,60){\rotatebox{90}{$\mathrm{Im}(z)$}}
\label{fig:U1_flow_configs_Tdep_LargeKappa}
}\end{minipage}\; \;
\begin{minipage}{.48\textwidth}
\subfloat[$\Lambda_F$ dependence at $T_{\mathrm{flow}}=1.5$]{
\includegraphics[scale=0.3]{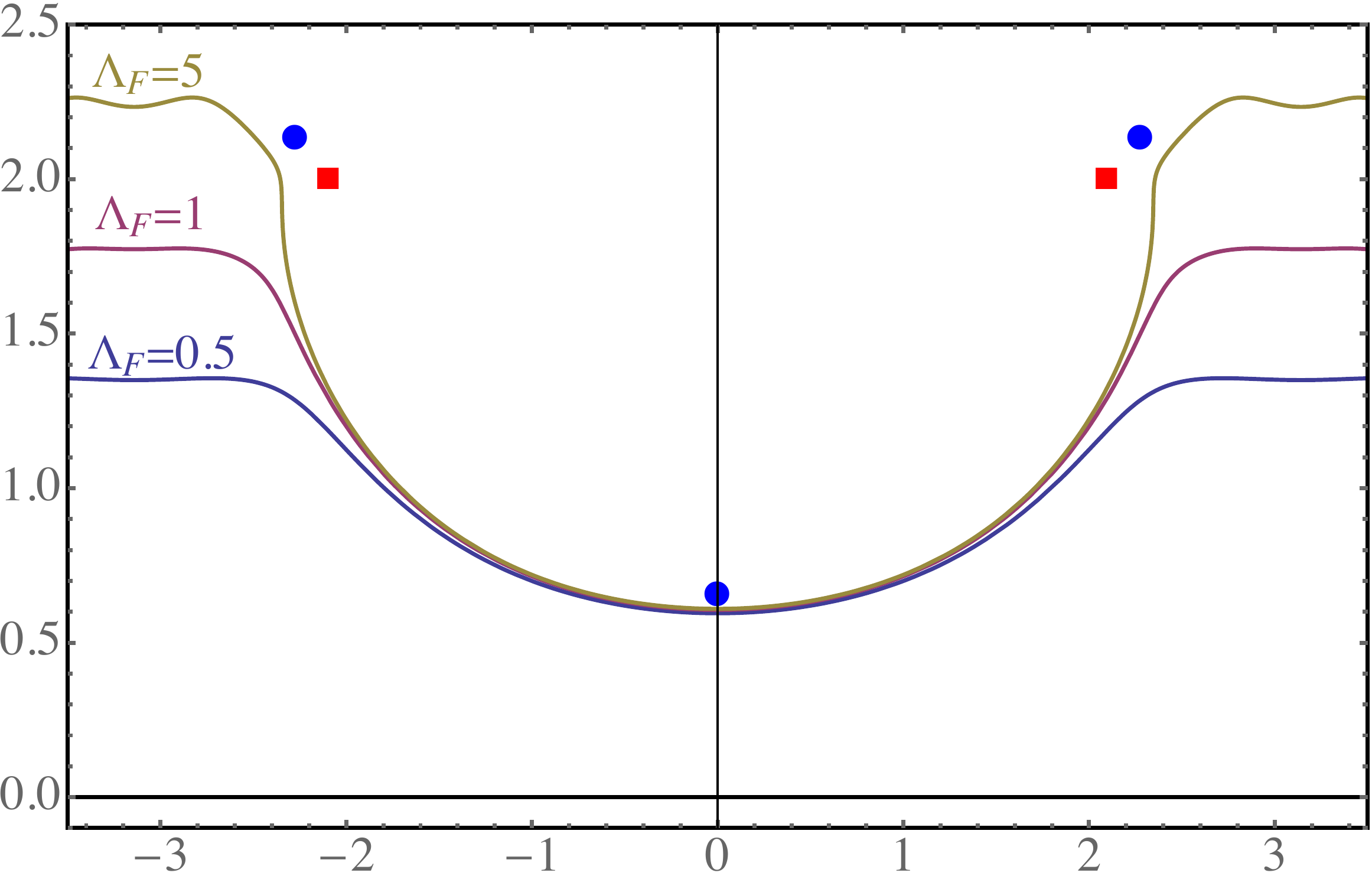}
\put(-110,-10){$\mathrm{Re}(z)$}
\put(-213,60){\rotatebox{90}{$\mathrm{Im}(z)$}}
\label{fig:U1_flow_configs_Ldep_LargeKappa}
}\end{minipage}
\caption{Complex contours $\mathcal{J}_{\Lambda}(T_{\mathrm{flow}})$ for the $U(1)$ one-link model with $\kappa=2$. We set $\Lambda_B=10$. The saddle points,
  which in the left figure are intersected by the Lefschetz thimble, are denoted by
the blue dots while the zero or the fermion determinant are shown as red squares. The dashed black curves show flow lines with initial condition on the real axis.}
\label{fig:U1_flow_config}
\end{figure}

We take $\kappa=2$, $\beta =1$, and $\mu =2$ in (\ref{eq:partition_function_U1}), and we set $\Lambda_B=10$ throughout the analysis in this case. The relevant saddle points are approximately given by 
\be
z_1\simeq 0.65\im,\quad z_{2,3}\simeq \pm2.28+2.13\im. 
\ee
The values of the classical action at these saddle points are $S_1\simeq -2.9$ and $S_{2,3}\simeq 3.8\pm5.7\im$, and thus the contribution is dominated by $z_1$. 
The zeros of the fermion determinant are given by 
\be
z_{*\pm}=\pm(\pi-\cos^{-1}(1/\kappa))+\im \mu. 
\ee
These zeros are shown with red squares in Fig.~\ref{fig:U1_flow_configs_Tdep_LargeKappa}, and the blue solid curves show the Lefschetz thimbles contributing to $Z$.

In Fig. ~\ref{fig:U1_flow_configs_Tdep_LargeKappa} we study the $T_{\mathrm{flow}}$-dependence of $\mathcal{J}_{\Lambda}(T_{\mathrm{flow}})$ at $\Lambda_F=5$. In this case, $\Lambda_B$ prevents the blow-up in the direction $z\to \pi+\im \infty$. 
The $\Lambda_F$-dependence of $\mathcal{J}_{\Lambda}(T_{\mathrm{flow}})$ is studied in Fig.~\ref{fig:U1_flow_configs_Ldep_LargeKappa}, and the contour becomes similar to Lefschetz thimbles as $\Lambda_F$ becomes larger. 

\begin{figure}[t]
\centering
\includegraphics[scale=0.4]{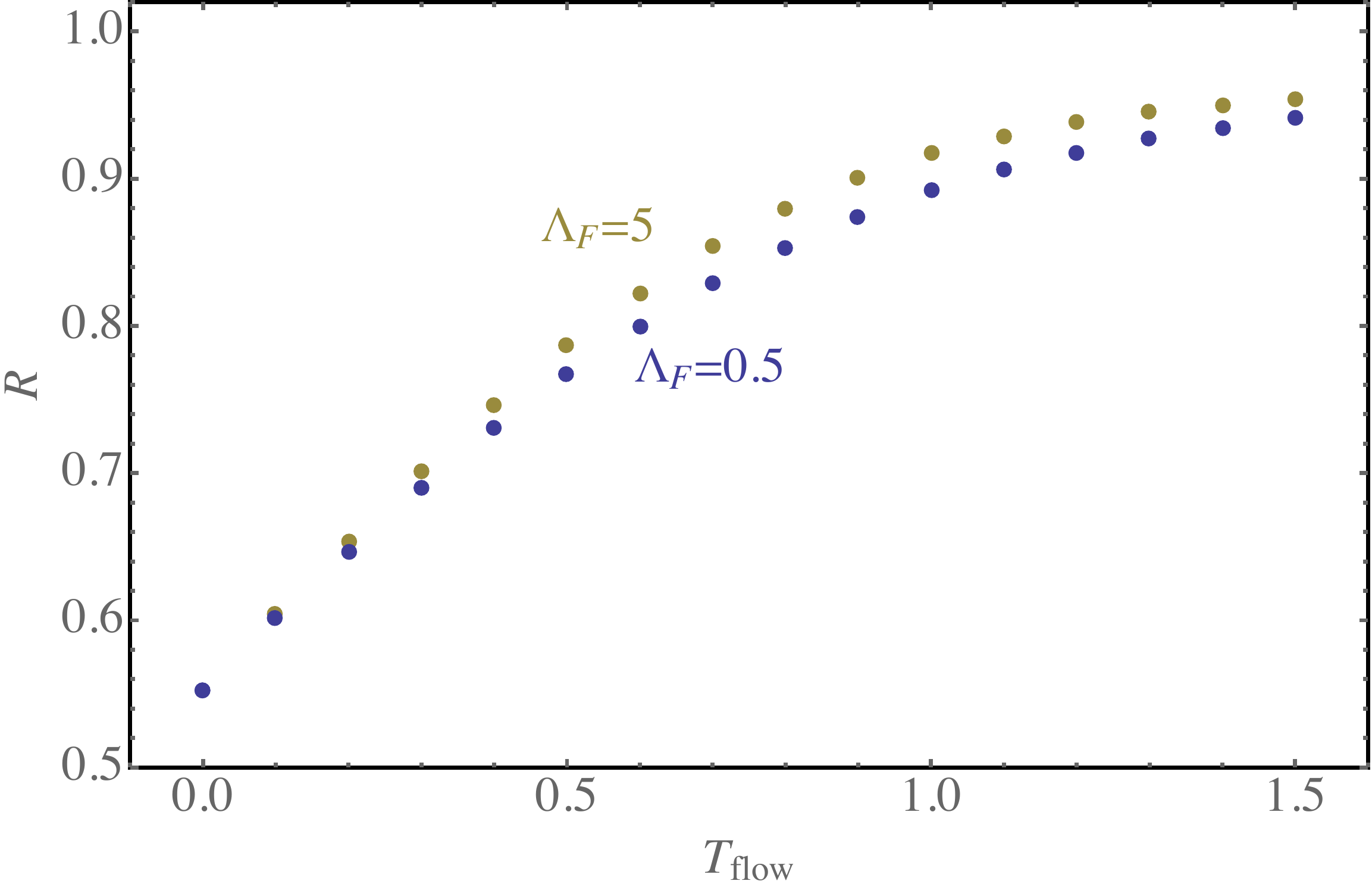}
\caption{$T_{\mathrm{flow}}$-dependence of the reweighting factor for the $U(1)$ one-link model with $\kappa=2$ and $\Lambda_B=10$. }
\label{fig:U1_reweighting_LargeKappa}
\end{figure}

Figure~\ref{fig:U1_reweighting_LargeKappa} shows the $T_{\mathrm{flow}}$-dependence of the reweighting factor at $\Lambda_F=0.5$ and $5$. The $\Lambda_F$-dependence of the reweighting factor is quite small. Within the time interval in our computation, the reweighting factor monotonically increases for this parameter. 
Compared with other examples studied in this paper, this model is an important benchmark because both $\Lambda_B$ and $\Lambda_F$ are effective to prevent the blow-up in this parameter region. We have checked through this model that the reweighting factor behaves as we have expected even in such situations. 

%-------------------------------------------------------------------------------%
\section{Conclusions}\label{sec:conclusion}

We have argued that the conventional gradient flow defining Lefschetz thimbles generically blows up, and thus one needs to monitor the divergence of gradient flows with great care in numerical computations. 
Instead of doing that, we propose a new gradient flow equation (\ref{eq:new_flow01}) that also defines Lefschetz thimbles and does not suffer from blow-ups. 
We show its theoretical foundation by providing a geometric interpretation
of the change in the gradient flows, and also prove rigorously that our new flow equation does not have blow-ups. 
In some examples of one-dimensional integrals with a sign problem, we  numerically construct the complex contours using the new gradient flow to see how it works in practice. 
By appropriately choosing the regularization parameters of the new gradient flow, we check that it solves the sign problem as the conventional flow equation does. 

One possible concern about our proposal would be the numerical cost, but we believe that it is not the problem for the following reasons. Since the computation of the metric needs the absolute value of the fermion determinant, it takes $O(N^3)$ in the LU decomposition when $N$ is the size of the fermion matrix. 
However, one needs to compute the inverse of the fermion matrix even without introducing the metric and it also costs $O(N^3)$, so this additional computational cost would not be a severe problem. 
Moreover, if the precise evaluation of the determinant is too costly, we could use the stochastic estimation of the determinant that reduces the cost significantly. 

Let us also briefly discuss another possible remedy treating the blow-up and compare it with our proposal. 
A simple remedy that uses the conventional flow equation would be the following: Introducing a cutoff in the process of
solving the conventional flow equation in order to estimate the blow-up, we throw away a trial configuration when it
satisfies a preset blow-up criterion. The argument is that blow-up occurs  when the action diverges and such configurations are suppressed anyway.
However, for strongly-coupled field theories, the configurations with exponentially small Boltzmann weights can give a significant contribution because of the exponentially large entropy, and we must check the cutoff-independence of the results obtained using this simple remedy. 
In our proposal, although additional costs are required to compute the metric, we do not introduce any cutoffs in doing the Monte Carlo algorithm. Since the additional computational costs for the metric is at most the same order of the computation for original flow equations, the check of cutoff-independence and our proposal would be comparable. 
Another possible merit for introducing the metric is that the flow equations becomes more stable than the original ones because of the absence of a blow-up, which allows to increase the step-size of the flow equation reducing  the computational cost. 

Our proposal (\ref{eq:new_flow01}) is only one possible way to introduce a metric in the gradient flow that prevents blow-up. We have shown that other choices define an equivalent Lefschetz-thimble decomposition so long as the gradient flow takes the form (\ref{eq:new_flow03}). 
We therefore would like to comment that technical problems of solving gradient flow equations might be circumvented by choosing a different metric. 

It would be interesting to see how our proposal works for the path integral
of more realistic systems with strong interaction. 
Toward the final goal of computing finite-density QCD, chiral random matrix is a good candidate to be tested. 
Indeed, previous studies~\cite{Gibbs:1986ut, Gibbs:1986hi, Barbour:1997bh, Barbour:1997ej, Stephanov:1996ki, Cohen:2003kd, Splittorff:2006fu, Splittorff:2007ck, Cherman:2010jj, Hidaka:2011jj, Nagata:2012tc} reveal that chiral symmetry breaking and associated charged pions are the origin of the difficulties for the
numerical simulations of cold and dense QCD. Since chiral random matrix theory shares the same universality with regards to the Dirac-eigenvalue distributions, its systematic study with Lefschetz thimbles, which is partly done in Ref.~\cite{DiRenzo:2015foa}, will provide us an important insight in this problem. 

%-----------------------------------------------------------------------------%
\acknowledgments
Y.~T. thanks Francesco Di Renzo and Giovanni Eruzzi for useful discussions and their kindness at VIII Parma International School of Theoretical Physics. 
Y.~T. and H.~N. are financially supported by Special Postdoctoral Researchers program of RIKEN. 
%Appendices--------------------------------------------------------------------%
\appendix

\section{Proof of equivalence among gradient flows}\label{app:proof_equivalence}
In this Appendix, 
we prove that the Lefschetz-thimble decompositions with different Hermitian metrics $g_{i\overline{j}}$ are equivalent. 
We assume that $S(z)$ is a polynomial on $\mathbb{C}^n$ without degenerate critical points and critical points at infinities. 
We further assume that the $\mathrm{Im}(S(z_{\sigma}))$ are all different
for different saddle points.

We denote the solution of the gradient flow with the metric $g$ as $z^{(g)}(t)$, and define the Lefschetz thimble and its dual by 
\bea
\mathcal{J}^{(g)}_{\sigma}&=&\{z^{(g)}(0)\in\mathbb{C}^n\,|\, z^{(g)}(-\infty)=z_{\sigma}\},\\ 
\mathcal{K}^{(g)}_{\sigma}&=&\{z^{(g)}(0)\in\mathbb{C}^n\,|\, z^{(g)}(+\infty)=z_{\sigma}\}. 
\eea
We can show that 
\be
\langle \mathcal{J}^{(g)}_{\sigma},\mathcal{K}^{(g)}_{\tau}\rangle =\delta_{\sigma\tau}. 
\ee
This is because $\mathrm{Im}(S(z))$ is constant on each $\mathcal{J}^{(g)}_{\sigma}$, $\mathcal{K}^{(g)}_{\tau}$, and hence the above assumption on $\mathrm{Im}(S)$ implies $\mathcal{J}^{(g)}_{\sigma}$ and $\mathcal{K}^{(g)}_{\tau}$ cannot intersect when $\sigma\not=\tau$. We should notice that $\mathcal{J}^{(g)}_{\sigma}$ and $\mathcal{K}^{(g)}_{\sigma}$ intersects only at $z_{\sigma}$ since $\mathrm{Re}(S(z))$ is monotonically increasing along the flow. 
Moreover, the same argument shows that 
\be
\langle \mathcal{J}^{(g)}_{\sigma},\mathcal{K}^{(g')}_{\tau}\rangle =\delta_{\sigma\tau} 
\ee
even when $\mathcal{J}$ and $\mathcal{K}$ are defined by different metrics $g$, $g'$. 

Since $\mathrm{Re}(S(z))$ increases  monotonically along the gradient flow, $\mathcal{J}^{(g)}$ defines integration cycles of the form $\mathrm{e}^{-S(z)}\diff^n z$. Therefore, 
\be
[\mathcal{J}^{(g)}_{\sigma}]\in H_n(\mathbb{C}^n,\{\mathrm{e}^{-\mathrm{Re}(S)}\ll 1\})\simeq \sum_{\sigma}\mathbb{Z}\, \mathcal{J}_{\sigma}. 
\ee
Similarly, 
\be
[\mathcal{K}^{(g)}_{\sigma}]\in H_n(\mathbb{C}^n,\{\mathrm{e}^{-\mathrm{Re}(S)}\gg 1\})\simeq \sum_{\sigma}\mathbb{Z}\, \mathcal{K}_{\sigma}. 
\ee
Using the above property on the intersection pairing, we obtain the identity for the homology class,
\be
[\mathcal{J}^{(g)}_{\sigma}]=\sum_{\tau}\langle \mathcal{J}^{(g)}_{\sigma},\mathcal{K}_{\tau}\rangle [\mathcal{J}_{\tau}]=[\mathcal{J}_{\sigma}]. 
\ee 
Similarly, we obtain 
\be
[\mathcal{K}^{(g)}_{\sigma}]=[\mathcal{K}_{\sigma}]. 
\ee
This shows that the homology class of the Lefschetz thimble does not depend on the choice of Hermitian metric. It also shows that 
\be
Z=\sum_{\sigma}\langle\mathbb{R}^n,\mathcal{K}_{\sigma}\rangle\int_{\mathcal{J}^{(g)}_{\sigma}}\diff^n z \exp\left[-S(z)\right] 
\ee
for any choice of $g$. Here, we emphasize again that the coefficients $\langle\mathbb{R}^n,\mathcal{K}_{\sigma}\rangle$ are independent of the choice of metric $g$. It partly comes from the fact that the intersection number is a topological quantity while the metric is a regular complex function. The intersection number thus jumps only when the Stokes phenomenon happens, and what we have shown here is that the change of the metric does not cause the Stokes jumping. 

There are a few remarks on this result. For the one-dimensional examples, the Lefschetz thimble is nothing but the stationary phase contour, that is characterized by $\mathrm{Im}(S(z))$ being constant. 
Since this amounts to one constraint in the two-dimensional space $\mathbb{C}\simeq\mathbb{R}^2$, the Lefschetz thimble is uniquely defined as a submanifold. Independence of the metric is trivial since the stationary phase condition does not use the metric. 
On the other hand, this is highly nontrivial if one considers higher-dimensional integrals. In this case, the stationary phase condition is insufficient to characterize the half-dimensional submanifolds in $\mathbb{C}^n$, and there are a lot of possible choices for steepest descent cycles. Therefore, $\mathcal{J}_{\sigma}$ and $\mathcal{J}^{(g)}_{\sigma}$ can be different submanifolds of $\mathbb{C}^n$ if $n\ge 2$. 
Application of the Picard--Lefschetz theory ensures that all of them are ``equivalent'' in the sense that their homology class is the same.

\section{Flow equation for the Jacobian matrix}\label{app:flow_jacobian}
For the numerical computation of the Lefschetz-thimble Monte Carlo method,
we do not only need the flow $z(T,x)$ but also the flow of the Jacobian $\mathrm{det}\left({\p z^i(T,x)/\p x^j}\right)$. 
We first derive the result for the most general expression (\ref{eq:new_flow03}), and then apply it to the cases (\ref{eq:new_flow01}) and (\ref{eq:new_flow02}). 

Let us consider two solutions with infinitesimally close initial conditions $z(T,x)$ and $z(T,x+\Delta x)$, where $|\Delta x|\ll 1$. 
We compute the deviation of the gradient flow (\ref{eq:new_flow03}) as 
\be
{\diff \Delta z^i\over \diff t}=\left(g^{i\overline{j}}\p_{\overline{j}}\p_{\overline{k}}\overline{S}+\p_{\overline{k}}g^{i\overline{j}}\p_{\overline{j}}\overline{S}\right)\overline{\Delta z^k}+\p_k g^{i\overline{j}}\p_{\overline{j}}\overline{S}\Delta z^k. 
\ee
Here, we introduced the shorthand notation $\p_i={\p/\p z^i}$ and $\p_{\overline{i}}=\p/\p\overline{z^i}$, and $\Delta z=z(t,x+\Delta x)-z(t,x)$. 
By writing the Jacobian matrix by 
\be
J^{i}_{j}(t,x)={\p z^i(t,x)\over \p x^j}, 
\ee
$\Delta z$ becomes $\Delta z^i=J^{i}_{j}\Delta x^j$. To compute the Jacobian, we consider a real-valued variation for $\Delta x$, and thus we obtain
\be
{\diff J^i_{\ell}\over \diff t}=\left(g^{i\overline{j}}\p_{\overline{j}}\p_{\overline{k}}\overline{S}+\p_{\overline{k}}g^{i\overline{j}}\p_{\overline{j}}\overline{S}\right)\overline{J^k_{\ell}}+\p_k g^{i\overline{j}}\p_{\overline{j}}\overline{S}J^k_{\ell}
\ee
by comparing coefficients of $\Delta x^\ell$. If one assumes that $g^{i\overline{j}}\propto \delta^{i\overline{j}}$ at the saddle points $\p_i S=0$, then one can solve this equation in the vicinity of a saddle point by applying Takagi's factorization to $\p_{\overline{i}}\p_{\overline{k}}\overline{S}$. 

Let us restrict ourselves to diagonal metrics $g^{i\overline{j}}(z,\overline{z})=g(z,\overline{z})\delta^{i\overline{j}}$. Then, we obtain 
\bea
{\diff J^i_{\ell}\over \diff t}&=&\left(g\p_{\overline{i}}\p_{\overline{k}}\overline{S}+\p_{\overline{i}}\overline{S}\p_{\overline{k}}g\right)\overline{J^k_{\ell}}+\p_{\overline{i}}\overline{S}\p_k gJ^k_{\ell}\nonumber\\
&=&g\left[\left(\p_{\overline{i}}\p_{\overline{k}}\overline{S}+\p_{\overline{i}}\overline{S}\p_{\overline{k}}\ln g\right)\overline{J^k_{\ell}}+\left(\p_{\overline{i}}\overline{S}\p_k \ln g\right) J^k_{\ell}\right]. 
\eea
For the conventional gradient flow (\ref{eq:conventional_flow}) , we reproduce the well-known formula~\cite{Cristoforetti:2012su}
\be
{\diff J^i_{\ell}\over \diff t}=\left(\p_{\overline{i}}\p_{\overline{k}}\overline{S}\right)\overline{J^k_{\ell}}
\ee
by substituting $g=1$. 

For the proposed gradient flow (\ref{eq:new_flow01}), we need to compute $\p_k \ln g$ for the metric (\ref{eq:metric_new_flow1}): 
\be
\p_k \ln g=-{\p_k S_B\over \Lambda_B}+{\Lambda_F^{-2}D^{-1}\p_k D\over |D|^2+\Lambda_F^{-2}}. 
\ee
We obtain $\p_{\overline{k}}\ln g$ by taking the complex conjugation since $g$ is real. As a result, the deviation equation for (\ref{eq:new_flow01}) is given by 
\bea
{\diff J^i_\ell\over \diff t}&=&\mathrm{e}^{-2\mathrm{Re}(S_B)/\Lambda_B}{|D|^2\over |D|^2+\Lambda_F^{-2}}\left[
\left\{\p_{\overline{i}}\p_{\overline{k}}\overline{S}+\p_{\overline{i}}\overline{S}\left(-{\p_{\overline{k}} \overline{S_B}\over \Lambda_B}+{\overline{D}^{-1}\p_{\overline{k}} \overline{D}\over \Lambda_F^2|D|^2+1}\right)\right\}\overline{J^k_{\ell}}\right.\nonumber\\
&&\left.\quad+\left\{\p_{\overline{j}}\overline{S}\left(-{\p_k S_B\over \Lambda_B}+{D^{-1}\p_k D\over \Lambda_F^2|D|^2+1}\right)\right\} J^k_{\ell}
\right]. 
\eea

\section{Comment on the Hermitian and K\"ahler metric in the gradient flow}\label{app:complex_str}

In the original applications of Lefschetz thimbles to quantum gauge theories~\cite{Witten:2010cx, Witten:2010zr}, the K\"ahler nature of the complexified field space was emphasized. Our proposal~(\ref{eq:new_flow01}), however, introduces the Hermitian metric in the gradient flow, and it is not K\"ahler. In this appendix, we will justify the use of our proposal even for the sign problem of lattice gauge theories. 

\subsection{Quick review on complex structure}
This is the brief summary of  Hermitian and K\"ahler structures. In the following, we consider a $2n$-dimensional smooth (real) manifold $M$. 

If there exists a bundle map $J:TM\to TM$ with $J^2=-1$ and $TM$ the tangent bundle of $M$, we call $J$ an almost complex structure on $M$. 
By considering a complexification of the tangent bundle $TM\otimes_{\mathbb{R}}\mathbb{C}$, one can diagonalize $J$ at each point $p\in M$ with eigenvalues $\pm\sqrt{-1}$ and degeneracies  $n_\pm$. 
If $J=J^{\nu}_{\mu}\p_{\nu}\otimes \diff x^{\mu}$ satisfies the integrability condition, 
\be
J^{\nu}_{\mu}(\p_{\rho}J^{\mu}_{\sigma}-\p_{\sigma} J^{\mu}_{\rho})+J^{\mu}_{\sigma}\p_{\mu}J^{\nu}_{\rho}-J^{\mu}_{\rho}\p_{\mu} J^{\nu}_{\sigma}=0, 
\ee
we say $J$ is a complex structure of $M$, and $M$ is called a complex manifold. 

If $M$ is a complex manifold, we can take local (holomorphic and anti-holomorphic) coordinates $z^{i}$ and $z^{\overline{i}}=\overline{z^i}$ ($i=1,\ldots,n$), which diagonalizes $J$ at each point and the transformation property among them is holomorphic, thanks to Newlander--Nirenberg theorem. More concretely, in such coordinates the complex structure looks like 
\be
J^{j}_{i}=\sqrt{-1}\delta^{j}_{i},\, J^{\overline{j}}_{\overline{i}}=-\sqrt{-1}\delta^{\overline{j}}_{\overline{i}},\, J^{\overline{j}}_{i}=J^{j}_{\overline{i}}=0. 
\ee
If one takes another coordinate patch $w^i$ and $w^{\overline{i}}$ with the same property, then $w^i(z)$ are holomorphic and $w^{\overline{i}}(\overline{z})$ are anti-holomorphic thanks to the integrability condition. 

\subsubsection{Hermitian structure}
A Riemannian manifold with an (almost) complex structure $(M,g,J)$ is called Hermitian if it satisfies 
\be
g_{\mu\nu}J^{\mu}_{\rho}J^{\nu}_{\sigma}=g_{\rho\sigma}. 
\ee
Let us take a holomorphic local coordinate $z^i$, then this condition implies
\be
(\diff s)^2 = g_{i\overline{j}}(\diff z^i \otimes \diff z^{\overline{j}}+\diff z^{\overline{j}}\otimes \diff z^i). 
\ee
That is, $g_{ij}=g_{\overline{i}\,\overline{j}}=0$. 
Using the mixed component of the metric, one can define a non-degenerate $2$-form, called the Hermitian form, 
\be
\omega=\im g_{i\overline{j}}\diff z^i \wedge \diff z^{\overline{j}}. 
\ee

\subsubsection{K\"ahler structure}
A Riemannian manifold with a complex structure $(M,g, J)$ is called K\"ahler if it satisfies 
\begin{itemize}
\item $D_\mu J^{\nu}_{\rho}=0$ ($D_{\mu}$ is the covariant derivative with the Levi-Civita connection), 
\item $g_{\mu\nu}J^{\mu}_{\rho}J^{\nu}_{\sigma}=g_{\rho \sigma}$. 
\end{itemize}
Compared to the Hermitian case, the first condition is imposed additionally  for the K\"ahler manifold. 
Let us take a holomorphic local coordinate $z^i$ again, then the second condition implies that $g_{ij}=g_{\overline{i}\, \overline{j}}=0$, i.e., 
\be
(\diff s)^2 = g_{i\overline{j}}(\diff z^i \otimes \diff z^{\overline{j}}+\diff z^{\overline{j}}\otimes \diff z^i). 
\ee
Using the mixed component of the metric, one can define a $2$-form, called the K\"ahler form, 
\be
\omega=\im g_{i\overline{j}}\diff z^i \wedge \diff z^{\overline{j}}. 
\ee
The first condition means that $\p_i g_{j\overline{k}}=\p_j g_{i\overline{k}}$ and $\p_{\overline{i}} g_{j\overline{k}}=\p_{\overline{k}} g_{j\overline{i}}$, and it is equivalent to say that $\omega$ is symplectic, i.e.,
\be
\diff \omega=0. 
\ee
That is, one can say that K\"ahler manifolds are Hermitian manifolds whose Hermitian forms are symplectic. 
This condition ensures the existence of a local function $K$, which is called a K\"ahler potential, so that $\omega=\im \p\overline{\p}K$, where $\p$ and $\overline{\p}$ are holomorphic and anti-holomorphic exterior derivatives (they are called  Dolbeault operators, and locally $\p=\diff z^i \p_{i}$, etc.). Note that $K$ is not necessarily a (globally defined) function, which is why $K$ is called ``potential". 

%This simplifies the Christoffel symbol and the Riemann tensor. Non-zero components of Christoffel symbol are 
%\be
%\Gamma^i_{\, jk}=g^{i\overline{\ell}}\p_j g_{k\overline{\ell}}
%\ee
%and complex conjugate. That is, mixed components always vanish. Similarly, non-zero components of the Riemann tensor are 
%\be
%{R_{\overline{i}j}}^{k}_{\,\ell}=\p_{\overline{i}}{{\Gamma_{j}}^{k}}_{\ell}. 
%\ee
%The Ricci $2$-form can be summarized to $R=-\im \p \overline{\p}\ln \det(g_{i\overline{j}})$. 

\subsection{Gradient flow and Hamilton equation of motion}

In this section, we first review why the K\"ahler nature of the complexified space is useful for analytic applications of the Lefschetz-thimble method to topological
gauge theories~\cite{Witten:2010cx, Witten:2010zr}. In the last paragraph, we argue that
practical applications do not require the K\"ahler property, and we will conclude that the Hermitian metric can be used to define the flow equation of lattice gauge theories when treating the sign problem. 

Let us pick up a holomorphic map $S:M\to \mathbb{C}$, and consider the flow equation,
\be
{\diff z^i\over \diff t}=g^{i\overline{j}}\p_{\overline{j}}\overline{S}. \label{eq:holomorphic_flow}
\ee 
This can be viewed from two perspectives,  the Hermitian or the K\"ahler manifold. From the Riemannian nature of $M$, this is the gradient flow with the height function $\mathrm{Re}(S)=(S+\overline{S})/2:M\to\mathbb{R}$, 
\be
{\diff x^{\mu}\over \diff t}=g^{\mu\nu}\p_{\nu}2\mathrm{Re}(S). 
\ee
One can easily check that in the holomorphic coordinate this goes back to the original equation (\ref{eq:holomorphic_flow}) using the Cauchy--Riemann condition. 
In order to get another perspective, we introduce a ``bracket" defined from the Hermitian or K\"ahler form $\omega$: 
\be
\{f,g\}=-\im g^{i \overline{j}} (\p_i f \p_{\overline{j}}g-\p_i g \p_{\overline{j}}f). 
\ee
If we consider the following equation with $\mathrm{Im}(S)=(S-\overline{S})/(2\im):M\to \mathbb{R}$,
\be
{\diff x^{\mu}\over \diff t}=\{x^{\mu},2\mathrm{Im}(S)\}, 
\ee
it again gives the original equation (\ref{eq:holomorphic_flow}). Since $\{f,f\}=0$ in general, this elucidates that $\mathrm{Im}(S)$ is conserved along the flow equation.   

The huge merit in choosing the K\"ahler metric comes from the fact that $\{,\}$ becomes the Poisson bracket for the K\"ahler form $\omega$, i.e. it satisfies the Jacobi identity $\{f,\{g,h\}\}+\{g,\{h,f\}\}+\{h,\{f,g\}\}=0$. Therefore, for the K\"ahler metric, the gradient flow has a classical mechanical interpretation~\cite{Witten:2010cx, Witten:2010zr}.

If $\mathrm{Re}(S)$ is a Morse function on $M$ and satisfies the Morse--Smale condition (i.e., the
critical points of $S$ are non-degenerate and with all different $\mathrm{Im}(S)$ at those critical points), then one can compute basis of  relative homologies $H_n(M,\{\mathrm{e}^{-\mathrm{Re}(S)}\ll 1\})$ and $H_n(M, \{\mathrm{e}^{-\mathrm{Re}(S)}\gg 1\})$ using the gradient flow, as we have seen in Appendix~\ref{app:proof_equivalence}. Those bases are called Lefschetz thimbles and dual thimbles, respectively.  There exists a natural pairing called the intersection pairing, and this is important for the decomposition of the middle-dimensional cycles in terms of Lefschetz thimbles. 

However, the Morse--Smale condition for $\mathrm{Re}(S)$ is not always satisfied in practical applications to physics. Especially for gauge theories, the set of critical points is usually degenerate due to the gauge symmetry. In this case, the above equivalence between the gradient flow and the Hamilton equation is very helpful by choosing the K\"ahler metric~\cite{Witten:2010cx, Witten:2010zr} (see also \cite{Tanizaki:2014tua, Kanazawa:2014qma}). 
Let us call the symmetry group $\mathcal{G}$, then one can construct Noether charges $Q$ for this symmetry. 
One can consider the reduced phase space by performing the symplectic reduction $Q^{-1}(0)/\mathcal{G}$ (also called Marsden--Weinstein reduction)~\cite{arnold_mechanics}, and define the Lefschetz-thimble decomposition in the reduced phase space. 
Using the Lefschetz thimbles and dual thimbles computed in $Q^{-1}(0)/\mathcal{G}$, one can construct correct half dimensional cycles in $M$ by considering group actions of the Noether charge $Q$ so that the intersection number is well-defined. 
From this argument, it turns out to be quite helpful to choose the K\"ahler metric instead of the Hermitian metric in order to prove the existence of the Lefschetz-thimble decomposition when the classical action $S$ has continuous symmetries. 

On the other hand, the two important properties, ${\diff \over \diff  t}\mathrm{Re}(S)\ge 0$ and ${\diff \over \diff t}\mathrm{Im}(S)=0$, are satisfied in general for Hermitian metrics. So long as one has a program to construct half-dimensional cycles using the flow, the Hermitian property is good enough to cure the sign problem. 
Equation (\ref{eq:practical_multi_thimbles}) provides such a method, and thus our choice of metric in (\ref{eq:new_flow01}) can be used also for theories with continuous symmetries, especially gauge theories. 

\bibliographystyle{utphys}
\bibliography{QFT,lefschetz,./ref}
%\bibliography{./Lef_thimble}
\end{document}